\documentclass[%
reprint,
 amsmath,amssymb,
 aps,
]{revtex4-2}

\usepackage{xcolor}
\usepackage{graphicx}
\usepackage{float}
\usepackage{siunitx}
\usepackage{makecell}
\usepackage{soul}
\usepackage[normalem]{ulem}   

\usepackage{hyperref}
\hypersetup{
    colorlinks=true,
    linkcolor=blue,
    citecolor=blue,
    filecolor=blue,
    urlcolor=blue
}

\begin{document}

\title{Monte-Carlo simulations of the capture and cooling of alkali-metal atoms by a supersonic helium jet}
\author{Jeremy Glick, William Huntington, Michael Borysow, Kevin Wen, Daniel Heinzen }
\affiliation{Department of Physics, 
 University of Texas at Austin, Austin, TX 78712
}%

\author{Jacek K\l{}os and E. Tiesinga}
\affiliation{Joint Quantum Institute, National Institute of Standards and Technology and University
of Maryland, Gaithersburg, MD 20899}

\date{\today}

\begin{abstract}
We present three-dimensional Monte-Carlo simulations of the
capture of 1000 K $^7$Li or 500 K $^{87}$Rb atoms by a continuous supersonic $^4$He jet and show that intense alkali-metal beams form with narrow transverse and longitudinal velocity distributions. The nozzle creating the $^4$He jet is held at approximately 4 K.
These conditions are similar to those in the cold $^7$Li source developed by some of us as described in [Phy. Rev. A \textbf{107}, 013302 (2023)]. 
The simulations use differential cross-sections obtained from quantum scattering calculations of $^7$Li or $^{87}$Rb atoms with $^4$He atoms for relative collision energies between $k\times 1$ mK to $k\times 3000$ K, where $k$ is the Boltzmann constant. For collision energies larger than $\approx k\times 4$ K the collisions favor forward scattering, deflecting the $^7$Li or $^{87}$Rb atoms by no more than a few degrees. From the simulations, we find that about 1\,\% of the lithium atoms are captured into the $^4$He jet, resulting in a lithium beam with a most probable velocity of about $210$ m/s and number densities on the order of $10^{8}$ cm$^{-3}$. Simulations predict narrow yet asymmetric velocity distributions which are verified by comparing to fluorescence measurements of the seeded $^7$Li atoms. We find agreement between simulated and experimentally measured seeded $^7$Li densities to be better than 50\,\% across a range of $^4$He flow rates.  We make predictions for capture efficiency and cooling of $^{87}$Rb by a supersonic $^4$He jet.
The capture efficiency for $^{87}$Rb is expected to be similar to $^7$Li.
\end{abstract}
\maketitle


\section{Introduction}

Supersonic jets of typically noble gases serve as a valuable tool for producing intense cold beams.
Seeding these inert sources with  other atoms and molecules can also create cold beams of these particles \cite{Marsden_1967_seededJets,Hillenkemp_2002_supersonic_expansions}. 
In fact, these systems have long been used for studies in molecular physics, fluid mechanics, and molecular spectroscopy \cite{Tarbutt_2002_YbF,Aggarwal_2021_supersonic_ablation,Miller_2004_fluid,David_2016_spectroscopy,Smalley_1977_optical_spectroscopy}.  Supersonic expansions of $^4$He can reach milliKelvin temperatures in the moving frame of the jet, due to the adiabatic expansion of the carrier gas \cite{Hagena_1987_condensation,Wang_1988_mK_He_jet,Huntington_2022_Cold_Atom_Source}
and thus seeded species can be cooled to similarly low temperatures.

Historically, seeding of non-condensible atoms or molecules generally occurs by mixing into the carrier gas well in advance of the jet nozzle. Condensible species can be seeded just before or immediately after the nozzle; this arrangement is known as a Smalley source \cite{Smalley_1977_optical_spectroscopy, Duncabn_2012_ClusterSources}. 
In Ref.~\cite{Huntington_2022_Cold_Atom_Source}, however, an experimental apparatus is described in which $^7$Li atoms are efficiently captured by a $^4$He jet after the jet has significantly expanded and has cooled to temperatures below 10 mK from its initial temperature of 4.2 K at the cryogenically cooled nozzle. Post-nozzle seeding then eliminates the heat load on the nozzle and condensation that may otherwise occur with a continuous lithium source and relatively small nozzle diameters. 
Modeling the capture and thermalization of the seeded species is essential for optimizing the source's performance. Over the years, researchers have examined pre-nozzle seeded jets \cite{DePaul_1993_model_seeded_jets,Schullian_2015_inelastic_collisions_DSMC,Marsden_1967_seededJets}; to our knowledge, however, modeling of the post-nozzle capture of an effusive beam into a supersonic jet has not been carried out.

Determining conditions for efficient post-nozzle seeding requires knowledge of the collisional properties between alkali-metal atoms and noble gas atoms. Such properties have been extensively studied over the past 60 years.  
For example, the collision-energy dependence of the elastic scattering cross-sections for collisions between $^7$Li and noble gas atoms were measured in the early 1970s~\cite{Ury:1972,Dehmer:1972}.
Diffusion coefficients describing the propagation of density gradients of trace amounts of alkali-metal atoms  in noble-gas buffer gasses have been computed as well as measured \cite{Marrero1972,Hamel1986,Medvedev:2018,Pouliot2021}.
Potential energy surfaces as well as collisional rates coefficients have been computed as well ~\cite{Partridge:2001,Bruhl:2001,Kerkines:2002}.
The single bound state for the $^7$Li$^4$He system, for example predicted in Ref.~\cite{Kleinekathofer1999}, was observed in 2013 \cite{Tariq2013}. Ultra-cold, $\mu$K and mK  samples of $^7$Li and $^{87}$Rb atoms are  used to measure pressure in the ultra-high vacuum (UHV) regime \cite{Fagnan2009, Yuan2013, Makhalov2016, Shen2020, Shen2021, Ehinger2022, Barker2022, Klos2023}.
These devices rely on  precise knowledge of thermalized elastic rate coefficients between the ultracold alkali-metal atoms and the ambient-temperature atomic and molecular gasses in the UHV vacuum.
To validate these devices known pressures of ambient temperature background gases, often noble gases, are deliberately added into the vacuum chamber.

To model the seeding process, it is crucial to understand the velocity-phase-space distribution of supersonic expansions. Models of supersonic expansions have been performed using various numerical methods as a means of finding
approximate solutions to the Boltzmann equation for the phase space density of the atomic species and molecules. Early research utilized the method of moments approach \cite{Toennies_1977_theory,Edwards_1966_theory,Hamel_1966_theory}  enabling studies of the decoupling  of  the transverse and longitudinal temperatures when the jet transitions from continuum or fluid-like flow  to free molecular flow. Modern simulations are often performed using computational fluid dynamics (CFD) solvers \cite{Kaushik_2015_CFD,Wilkes_2006_CFD} or direct-simulation Monte-Carlo (DSMC) simulations \cite{Usami_1999_DSMC,Even_2015_DSMC,Schullian_2022_3D_DSMC}. Both CFD and DSMC can model turbulent flow and shock front formation as well as evaluate the role of nozzle geometry on beam brightness.

CFD solvers rely on finding numerical solutions to the Navier-Stokes or related differential equations.  DSMC methods, pioneered by G. Bird \cite{Bird_1994_DSMC}, are often favored for rarefied gas flows. Particles move in classical straight-line paths interspersed by particle-particle collisions that change the directions and velocities of atom or molecular pairs based on sampling from differential scattering cross-sections at the relevant relative kinetic energy. Macroscopic gas quantities are then determined by statistical averages. Accurate differential scattering cross-sections, however, are not always available for every molecular system and phenomenological cross-section models must be employed.

Here, we report  on three-dimensional (3D) Monte Carlo simulations of the capture  of $^7$Li and $^{87}$Rb by a supersonic helium jet. We quantify the fraction of lithium and rubidium atoms that become entrained in the jet over a wide range of helium flow rates using accurate theoretical differential cross-sections for $^7$Li-$^4$He and $^{87}$Rb-$^4$He scattering. Simulated density and velocity profiles are presented and analyzed. We compare data from $^7$Li experiments to validate our approach.

\section{Experimental Setup and Theoretical Treatment} 

\begin{figure}[t]
    \includegraphics[width=0.97\columnwidth,trim=10 0 0 0,clip]{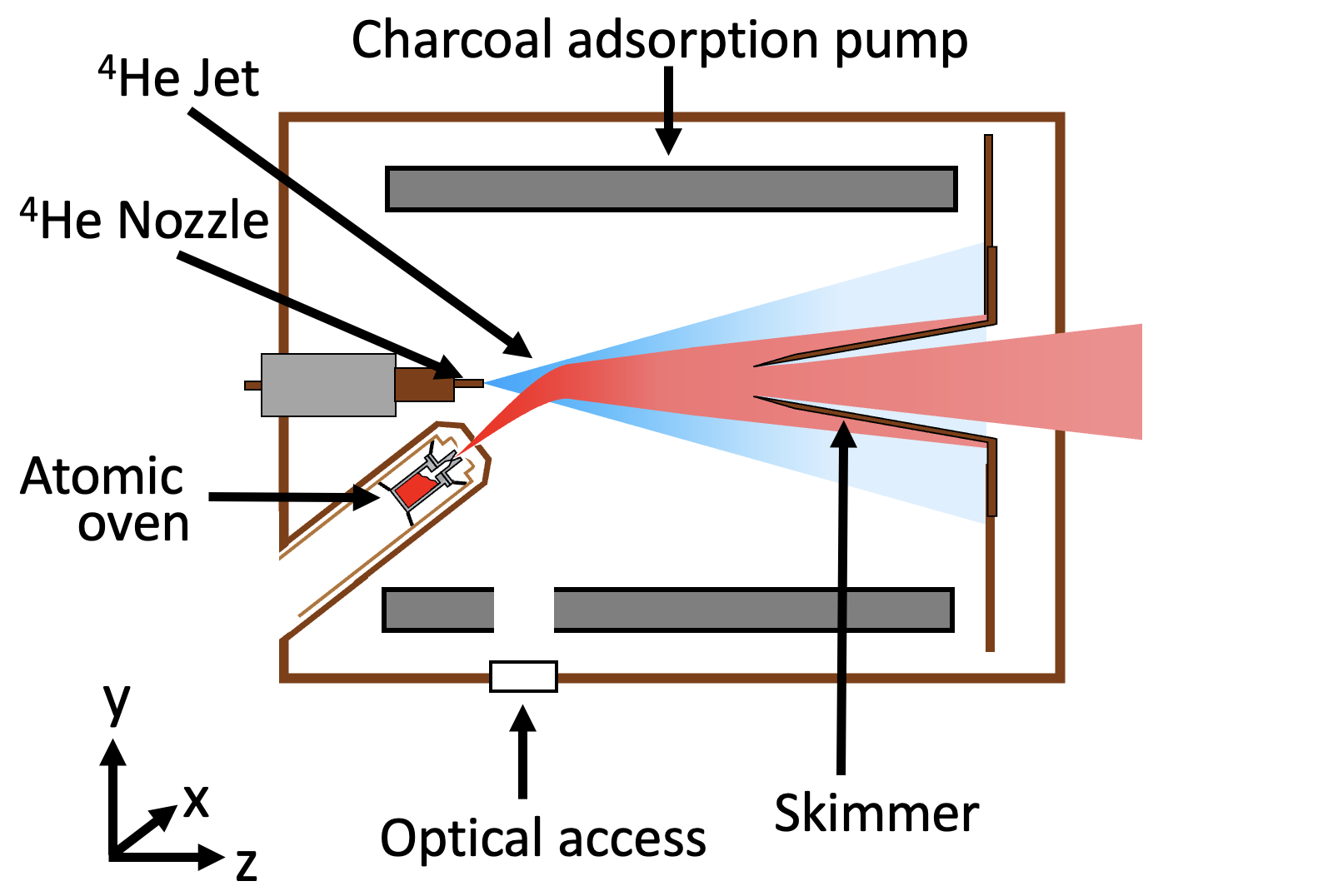}
    \caption{Schematic of the simulated apparatus showing lithium or rubidium atoms (red region) created in a hot atomic oven, captured by a supersonic cryogenic helium jet (cyan region), and then extracted as cold atoms with a skimmer.  The lower left corner shows the Cartesian $xyz$ coordinate system used in the simulations. The $y$ and $z$ axes are in the plane of this image, while the origin is at the tip of the cryogenic $^4$He sonic nozzle. System dimensions and helium jet and oven parameters can be found in the text.
    In the experimental realization, a charcoal adsorption pump is used to remove background helium to avoid the formation of shock fronts. A camera collects fluorescence emitted in the x-direction that is induced by a probe laser beam. The probe laser can be directed either vertically through the window labeled ``optical access" or horizontally along the center-line of the helium jet.}
    \label{apparatus}
\end{figure}

We model the dynamics of seeding and cooling energetic $^7$Li or $^{87}$Rb atoms into a supersonic helium jet. A schematic of the simulated apparatus and the definition of our coordinate system are shown in Fig.~\ref{apparatus}. The helium sonic nozzle is directed along the $(0,0,1)$ or $\hat z$ direction with its tip located at the origin $\vec x_{\rm N}=(0,0,0)$ of our coordinate system. The helium nozzle diameter $d_{\rm N}=0.020$ cm. Here and elsewhere system parameters
are motivated by the designs of Ref.~\cite{Huntington_2022_Cold_Atom_Source}.
The alkali-metal source creating an effusive beam of atoms is oriented along the  $(0,\sin\theta_{\rm S},\cos\theta_{\rm S})$ direction with angle $\theta_{\rm S}= 50^\circ$ from the $\hat z$ axis,
where the opening or aperture in front of the oven is located at $\vec x_{\rm S}=(0,-5.6,-3.0)$ cm. This source has a conical nozzle with an aperture diameter of $d_{\rm S} = 0.10$ cm.
The nozzle diameters are thus much smaller than $|\vec x_{\rm N}-\vec x_{\rm S}|$. Moreover,
the geometrical intersect or seeding distance between  the helium jet and  alkali-metal beam directions is 1.7 cm downstream from the helium nozzle. 
A skimmer extracts the cold alkali-metal atoms. The  skimmer is located 16 cm from the helium nozzle and has a circular opening with a diameter of 2.54 cm. The line connecting the  nozzle and the opening of the skimmer defines the center line of the helium jet. 

The orientation and location of the alkali-metal source relative to the helium jet are by no means unique.
For specificity, however, our simulations closely match the physical design constraints of the implementation in Ref.~\cite{Huntington_2022_Cold_Atom_Source}.
In principle, this restricts our ability to optimize the design. In practice, we find that
optimizing the flow rate of the $^4$He jet is sufficient for creating intense cold alkali-metal atom beams.

\subsection{Static helium jet}

The motion of all helium and alkali-metal atoms can, in principle, be modeled with the DSMC method. The efficient capture of the alkali-metal (AM) atoms, however, requires a dense helium jet. 
Our expected stagnation helium number densities, $n_0$, are a few times $10^{19}$ cm$^{-3}$ with a stagnation temperature of the helium atoms of about $T_0=4.2$ K. The stagnation conditions are the conditions in the reservoir before the nozzle. Based on these conditions, the mean free path between $^4$He-$^4$He collisions is a few microns using the $^4$He-$^4$He cross-section from Ref.~\cite{Chrysos_2017_He_He_cross_section}. This small mean free path puts the expansion well into the continuum regime and simulating the helium expansion directly with a DSMC method is challenging. Furthermore, we cannot take advantage of the two-dimensional (2D) axial symmetry of the helium jet and a three-dimensional (3D) simulation is required to model the injected alkali-metal atom beam.

We therefore introduce approximations based on the realization that the number densities of the alkali-metal atoms are orders of magnitude smaller than those in the helium jet. 
Firstly, we realize that the mean-free path between AM-AM collisions is larger than the centimeter size of the system and we only need to follow one alkali-metal atom at a time sampled from their phase-space density at the aperture of the atomic oven.
Secondly, the value for the mean-free path between AM-$^4$He collisions lies
between those for $^4$He-$^4$He and AM-AM collisions. Then the number density of the 
$^4$He jet and the relationships among the mean free paths have several implications.
The heat added to the helium jet as the alkali-metal atoms slow can be neglected. Consequently, we can assume that the helium jet is expanding adiabatically, is in the collisional regime, and is locally always in thermal equilibrium.
In other words, we can treat the jet as a static background flow and only need to determine lithium trajectories as they collide with $^4$He atoms. 

From Refs.~\cite{Pauly_2000_beams_book,miller_1988_free_jet}, it can be shown that the unit-normalized position and velocity phase-space probability distribution of the axially symmetric expansion of the helium atoms in the jet starting from position a few nozzle diameters away is well described by
\begin{eqnarray}
 \lefteqn{ P_{\rm He}(\vec r,\vec v)=}   \label{eq:PHelium}\\
 && \quad c_0 \left(\frac{m_{\rm He}}{2\pi k T_0}\right)^{3/2}\zeta(\hat r)   \exp\left[ - \frac{1}{2}\frac{ m_{\rm He} (\vec v -  v_{\rm He, tv}\, \hat r )^2}{kT_{\rm He}(r)}\right] \,,
  \nonumber
\end{eqnarray}
where $\hat r=\vec r/r$ is the orientation of $\vec r$ in  the Cartesian coordinate system
defined in Fig.~\ref{apparatus}. In spherical coordinates $\hat r$ is specified by polar angle $\theta\in[0,\pi]$ and azimuthal angle $\varphi\in[0,2\pi]$. 
In addition, the position-dependent helium temperature and radial terminal or streamline velocity are, 
\begin{equation}
   T_{\rm He}(r)= c_1\,T_0\, \left(\frac{d_{\rm N}}{r}\right)^{4/3}
   \label{eq:Thelium}
\end{equation}
and
\begin{equation}
   v_{\rm He, tv} = \sqrt{5\frac{k T_0}{m_{\rm He}}}\,,
   \label{eq:vtv}
\end{equation}
where $m_{\rm He}$ is the mass of the $^4$He atom and $k$ is the Boltzmann constant. Here, coefficients $c_0=1.62$ and $c_1=0.287$.
Finally, the angular function $\zeta(\hat r)$ is given by
\begin{equation}
   \zeta_{\rm He}(\hat r)= \frac{1}{{\cal Z}(t_{\rm He})}\cos^2(t_{\rm He} \theta) \,,
   \label{eq:fangular}
\end{equation}
for $0\le\theta <\pi/(2t_{\rm He})$ and zero otherwise. We have ${t_{\rm He}>1}$ and  ${\cal Z}(t_{\rm He})$ is defined such that 
$\int_0^\pi \sin\theta {\rm d}\theta \int_0^{2\pi} {\rm d}\varphi \,\zeta_{\rm He}(\hat r)=1$.  For our jet  $t_{\rm He}=1.15$ and $\pi/(2t_{\rm He}) = 78^\circ$.
Locally, after integrating $P_{\rm He}(\vec r,\vec v)$ over all velocities and using Eq.~(\ref{eq:Thelium}), we find that 
\begin{equation}
  p_{\rm He}(\vec r)=\int {\rm d}^3\vec v\, P_{\rm He}(\vec r,\vec v)
= c_0 c_1^{3/2} \left(\frac{d_{\rm N}}{r}\right)^2 \zeta_{\rm He}(\hat r) \,,
\label{eq:pdensity}
\end{equation}
and, similarly, that the root-mean-square velocity in the moving frame of the helium gas is
\begin{equation}
   v_{\rm He, rms}(\vec r)=  \sqrt{3\frac{k T_{\rm He}(r)}{m_{\rm He}}} \,.
\end{equation}

The  helium number density  is
\begin{equation}
     n_{\rm He}(\vec r) = n_0 p_{\rm He}(\vec r)=c_0 c_1^{3/2} n_0 \left( \frac{d_{\rm N}}{r}\right)^2\, \zeta_{\rm He}(\hat r),
     \label{eq:HeDensity}
\end{equation}

where 
\begin{equation}
n_0 = 0.513{\sqrt{\frac{1}{2}\frac{m_{\rm He}}{kT_0}}}\left( \frac{4}{\pi d_{\rm N}^2}\right)\dot{N}_{0} 
\label{eq:n0He}
\end{equation} 
and $\dot{N}_{0}$ are the helium number density and  number flow rate at the sonic nozzle, respectively \cite{miller_1988_free_jet}. 
Typical flow rates are between $2\times10^{19}$ atoms/s and $1\times10^{20}$ atoms/s \cite{Huntington_2022_Cold_Atom_Source}. 

The phase-space distribution defined in Eqs.~(\ref{eq:PHelium}) through (\ref{eq:fangular}), as stated before, breaks down for radii smaller than a few nozzle diameters. That is, the local helium temperature $T_{\rm He}(r)$  should approach $T_0$ for small $r$.
Equation (\ref{eq:PHelium}) also breaks down for large radii $r$, where the $^4$He-$^4$He collision rate is insufficient to maintain local thermal equilibrium.

In the experiments of Ref.~\cite{Huntington_2022_Cold_Atom_Source} and in our 
simulations, $T_0=4.2$ K and the largest allowed radius $r$ 
for which Eq.~(\ref{eq:PHelium}) is expected to be valid is about 10 cm for the lowest helium reservoir densities which we consider. From Eq.~(\ref{eq:Thelium}), the initial helium temperature implies that the helium temperature drops to $7$ mK approximately 1 cm away from the sonic nozzle.
The terminal velocity for $^4$He atoms is $\approx 210$ m/s. The terminal kinetic energy is $m_{\rm He} v_{\rm He, tv}^2/2= 5kT_0/2 \approx k\times 11$ K.  
As a general feature of supersonic expansion, we realize that $v_{\rm He, tv} \gg v_{\rm He, rms}(r)$
for $r\gg d_{\rm N}$.

The form for $\zeta_{\rm He}(\hat r)$ in Eq.~(\ref{eq:fangular}) is based on findings from early measurements in low-density wind tunnels \cite{Ashkenas_1965_density} as well as, more recently,  in supersonic molecular jets using linear Raman Spectroscopy \cite{Tejeda_1996_density}. Strictly speaking, the form was only observed for $\theta<20^\circ$ as shock fronts occurred in these experiments. 
In our model, we can assume that the background pressure in the apparatus is sufficiently low that the helium smoothly transitions to free molecular flow without the formation of shocks. 
In the experimental apparatus, a charcoal adsorption pump is used to remove background helium atoms and minimize the possibility of shock waves forming.

We have run simulations with other functional forms for $\zeta(\hat r)$, including the more complex profiles from Refs.~\cite{Ashkenas_1965_density, Tejeda_1996_density}. They differ mainly  for $\theta>20^\circ$ and we found negligible differences in capture rates of the alkali-metal atoms,  as few AM-$^4$He collisions occur for large angles $\theta$.
The explanation is that AM-$^4$He collisions in the periphery of the helium expansion are early on in the alkali-metal-atom slowing process so that collisions occur at large, $\sim k\times 100$ K collision energies nearly independent of the low helium jet temperature. 
Moreover, relatively few collisions occur in the periphery  due to the low helium density and small AM-$^4$He cross sections at large collision energies.

Finally, we note that in more accurate descriptions of the helium jet the transverse, $x$ and $y$, and longitudinal, $z$, temperature of the helium atoms need not be the same \cite{Toennies_1977_theory,Beijerinck_1981_perp_temp}.
The effects of orientation-dependent helium temperatures should be negligible for alkali-metal atoms entering the jet as the relative velocity is dominated by the alkali-metal-atom velocity. It might become important once the alkali-metal-atom's velocity becomes comparable to the local helium velocity. This, however, only occurs once an alkali-metal atom is already captured in the jet and does not affect the capture efficiency.
We have not included orientation-dependent helium temperatures in our simulations. 

\subsection{Collision Cross-Sections}

The Monte-Carlo simulations rely on an accurate knowledge of the differential cross-section 
\begin{equation}
\frac{{\rm d}\sigma_{\rm AM-He}}{{\rm d}\Omega} \,,
\end{equation} 
for  collisions of $^7$Li and $^{87}$Rb  with $^4$He, each in its $^2$S or $^1$S electronic ground state, as a function of relative collision energy $E$ and polar and azimuthal collision angles $\theta_{\rm c}$ and $\varphi_{\rm c}$. We will also use
the total elastic cross-section
\begin{equation}
\sigma_{\rm AM-He}(E)=\int_0^\pi\sin\theta_{\rm c} {\rm d}\theta_{\rm c}\int_0^{2\pi}{\rm d}\varphi_{\rm c}
\frac{{\rm d}\sigma_{\rm AM-He}}{{\rm d}\Omega} \,.
\end{equation}
Formal discussions of scattering theory describing cross-sections and  collision angles
but also partial waves, centrifugal barriers, and Wigner threshold laws
can be found in Refs.~\cite{Taylor,Child,Messiah}.

We rely on the recent evaluations of the relevant X$^2\Sigma^+$ potential energy curves as functions of  the interatomic separation $R$ defined as a distance between point-like atomic nuclei and numerical solutions of the Schr\"odinger equation for the relative motion of the atoms using these isotropic potentials by some of us in Refs.~\cite{Makrides2020,Makrides2022Ea,Klos2023}. 
An isotropic potential only depends on $R$ and not the orientation of the interatomic axis.
We use the reduced mass $\mu=m_{\rm AM} m_{\rm He}/(m_{\rm AM}+m_{\rm He})$ computed from atomic masses in the relative kinetic energy operator. 

In the simulation, the atomic oven produces $^7$Li or $^{87}$Rb atoms at temperatures of several hundreds of Kelvin
and no external magnetic field is applied.
Hence, we can assume that the eight hyperfine states of the electronic ground state of $^7$Li and $^{87}$Rb are 
equally populated. For our purposes, we can also ignore the $R$-dependence of the hyperfine energies of the alkali-metal atoms. Hence, the differential cross-sections are the same for all hyperfine states
of the alkali-metal atom.

Figure \ref{fig:dsdt1} shows ${\rm d}\sigma_{\rm AM-He}/{\rm d}\Omega$ as functions of collision angle $\theta_{\rm c}$ up to $10^\circ$ for selected collision energies between $E/k=5$ mK and 500 K. {Panel (a) of the figure shows data for the $^7$Li-$^4$He collision, while panel (b) shows data for $^{87}$Rb-$^4$He. For an isotropic potential, the differential cross section is independent of $\varphi_{\rm c}$. The relative uncertainty of differential cross-sections is below 5\,\% for $E/k>40$ K. For $E/k < 0.1$ K it is larger than 20\,\%.

We observed that for the largest collision energies shown in Fig.~\ref{fig:dsdt1} the differential cross-section drops off rapidly
with angle $\theta_{\rm c}$. For example, for $E\geq k\times 10$ K the half width at half maximum is less than a few degrees. For much smaller collision energies, the differential cross-section is independent of $\theta_{\rm c}$.
In other words, during the initial cooling process the alkali-metal atoms undergo only small deflections when colliding with helium atoms.

The angular dependence of ${\rm d}\sigma_{\rm AM-He}/{\rm d}\Omega$ can mostly be understood from the long-range dispersion or van-der-Waals  $-C_6/R^6$ behavior of the X$^2\Sigma^+$ potential, where the positive $C_6$ is the van-der-Waals dispersion coefficient. The dispersion interaction introduces natural length and energy scales. These are the van-der-Waals length $x_6=\sqrt[4]{2\mu C_6/\hbar^2}$ and van-der-Waals energy $E_6=\hbar^2/(2\mu x_6^2)$. With the dispersion coefficients from Ref.~\cite{DEREVIANKO2010323}, we find $x_6=21.4a_0$ and $28.0a_0$
for $^7$Li+$^4$He and $^{87}$Rb+$^4$He, respectively, where $a_0=0.0529177$ nm is the Bohr radius. Similarly, $E_6/k=74$ mK and 29 mK, respectively.

\begin{figure}
    \centering
    \includegraphics[scale=0.33, trim=0 0 0 0,clip]{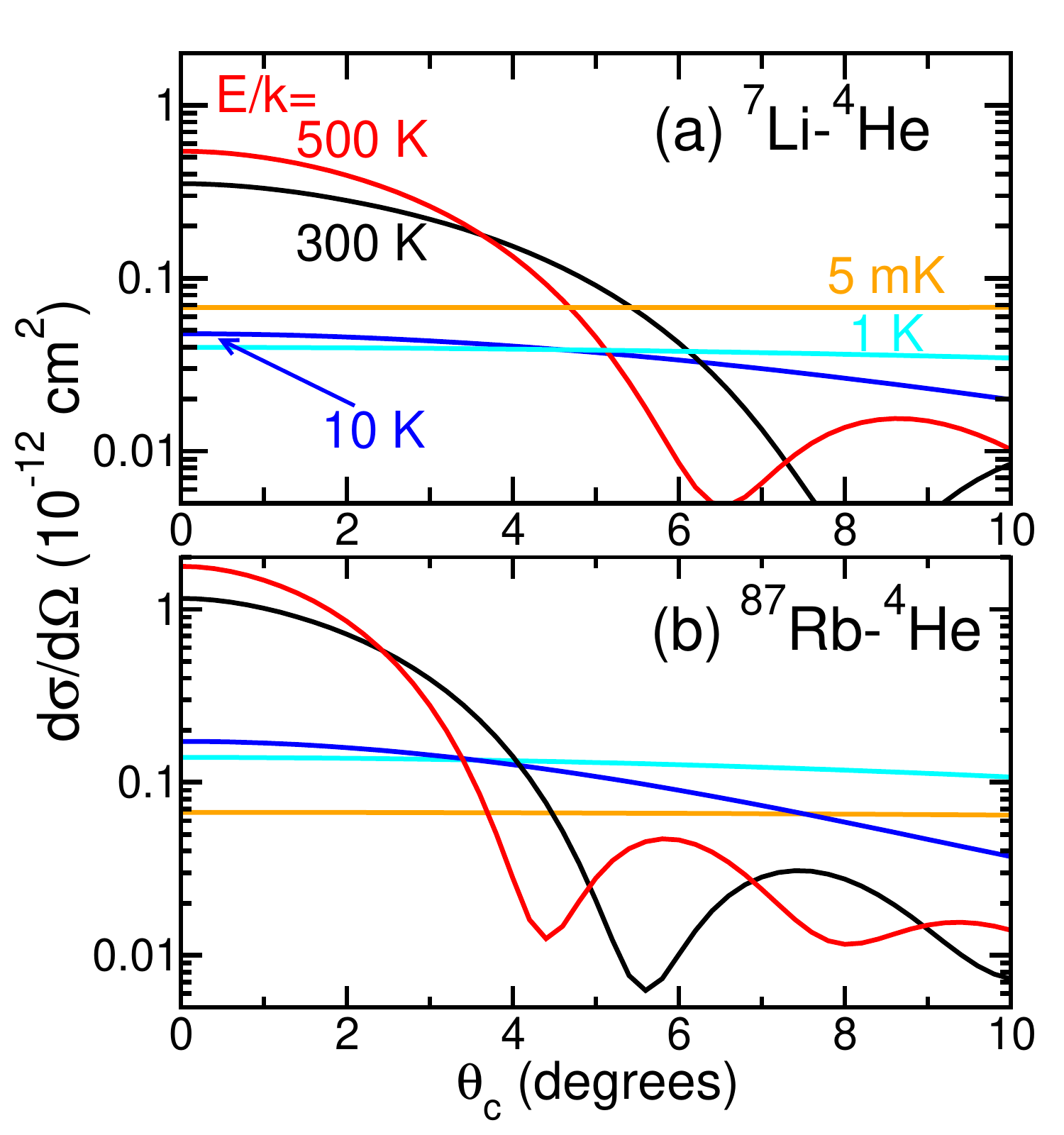}
    \caption{Differential cross sections ${\rm d}\sigma_{\rm AM-He}/{\rm d}\Omega$ for (a) $^7$Li+$^4$He and (b) $^{87}$Rb+$^4$He as functions of collision angle $\theta_{\rm c}$ for five relative kinetic energies $E$. The collision energies for the colored curves are the same in the two panels.}
    \label{fig:dsdt1}
\end{figure}

For  collision energies $E\ll E_6$, only a single partial wave, the so-called $s$ wave,  contributes to ${\rm d}\sigma_{\rm AM-He}/{\rm d}\Omega$ and we enter the Wigner threshold regime. Then ${\rm d}\sigma_{\rm AM-He}/{\rm d}\Omega\to a^2$ is independent of $\theta_{\rm c}$, where $a$ is the scattering length. 
For  collision energies $E\gg E_6$, a semi-classical approximation \cite{Child} gives a qualitative expression for the differential cross-section at small scattering angles $\theta_{\rm c}\ll\pi$. In fact, we have
\begin{equation}   
\frac{{\rm d}\sigma_{\rm AM-He}}{{\rm d}\Omega}= f_0 (E/E_6)^{3/5} \left\{ 1-f_1\, (E/E_6)^{4/5}(\theta_{\rm c}/2)^2 \right\} x_6^2 \,,
\end{equation}
with $f_0 = 0.363\,046\cdots$ and $f_1 =  2.018\,179\cdots$ corresponding to a sharply peaked differential cross-section. Still, a comparison with the quantum results in
Fig.~\ref{fig:dsdt1}, not shown, finds that the semi-classical model overestimates  ${\rm d}\sigma_{\rm AM-He}/{\rm d}\Omega$ at $\theta_{\rm c}=0$ by a factor two, while  it underestimates the half width at half maximum by a similar factor.

\subsection{Monte Carlo Simulation of the cooling of \texorpdfstring{$^7$}{7}Li or \texorpdfstring{$^{87}$}{87}Rb atoms}

The Monte-Carlo simulations begin by generating a $^7$Li or $^{87}$Rb particle at the center of the aperture in front of the alkali-metal source (AMS). 
The particle's velocity $\vec v_{\rm AM}$ is sampled from a Maxwell-Boltzmann distribution for an effusive source at temperature $T_{\rm AM}$ \cite{Greenland_1985_effusive_source, Pauly_2000_beams_book} with  probability distribution for its speed $v_{\rm AM}=|\vec v_{\rm AM}|$ given by
\begin{equation}
f_{\rm Li}(v) = \frac{1}{2}\left(\frac{m_{\rm AM}}{kT_{\rm AM}}\right)^2 v^3 \exp \left[-\frac{1}{2}\frac{m_{\rm AM}v^2}{kT_{\rm AM}} \right] \,,
\label{MaxBoltz_effusive}
\end{equation}
with $\int_0^\infty {\rm d}v \, f_{\rm Li}(v) =1$ and angular-velocity probability distribution given by $\xi(\theta,\varphi)=\cos(\theta)/{\cal N}(t_{\rm AMS})$ for $0\le\theta\le \pi/(2t_{\rm AMS})$ and zero otherwise with respect to the $(0,\sin\theta_{\rm S},\cos\theta_{\rm S})$ direction.
Here, $m_{\rm AM}$ is the mass of the alkali-metal atom and  ${\cal N}(t_{\rm AMS})$ is defined such that $\int_0^\pi \sin\theta {\rm d}\theta \int_0^{2\pi} {\rm d}\varphi \,\xi(\theta,\varphi)=1$. 
For our alkali-metal source $t_{\rm AMS}=\pi/0.18$ or $2t_{\rm AMS}/\pi=0.09$.
As the diameter of the aperture of the atomic source is small compared to the distance an alkali-metal atom travels before it enters the $^4$He jet, we can omit sampling over the initial location of the alkali-metal atom.

Once the initial position and velocity of the  $^7$Li or $^{87}$Rb atom is generated, the particle is propagated through the helium jet with variable time steps $\Delta t$ that are small fractions of the local mean-free time for collisions with $^4$He atoms. 
For much of the slowing process, the alkali-metal-atom's kinetic energy is large compared to $m_{\rm He}v_{\rm He,tv}^2/2$ and $kT_{\rm He}(r)$ of the helium atoms. 
This allows us to estimate the local mean-free time for an alkali-metal atom at position $\vec r_{\rm AM}$ and velocity $\vec v_{\rm AM}$ in this regime without the need for sampling the thermal distribution of the helium atoms and computing the local rate coefficient $K(\vec r_{\rm AM},\vec v_{\rm AM})=\langle v_{\rm rel} \sigma_{\textrm{AM--He}}(E)\rangle$, where the brackets indicate an average over the velocity distribution of the $^4$He atoms only, the relative velocity $\vec{v}_{\rm rel} = \vec{v}_{\rm AM} - \vec{v}_{\rm He}$, and $E=\mu v_{\rm rel}^2/2$.
The  mean free time is then well approximated by 
\begin{equation}
    \tau(\vec r_{\rm AM},\vec v_{\rm AM}) \approx \frac{1}{n_{\rm He}(\vec r_{\rm AM} ) v_{\rm approx} \,\sigma_{\textrm{AM--He}}(E_{\rm approx})}\,,
    \label{approx_gamma}
\end{equation}
where
\begin{equation}
    \label{rel_velocity}
    v_{\rm approx}= |\vec{v}_{\rm AM}-v_{\rm He, tv}\hat r_{\rm AM} |\,,
\end{equation}
and $E_{\rm approx}=\mu v_{\rm approx}^2/2$.
Once the velocity of an alkali-metal atom approaches the $\approx 210$ m/s terminal velocity of the helium atoms, Eq.~(\ref{approx_gamma}) loses accuracy. Hence, for $|\vec{v}_{\rm AM}-v_{\rm He, tv}\hat r_{\rm AM} |< v_{\rm cutoff}$, where $v_{\rm cutoff}=50$ m/s, the mean-free time at  $\vec r_{\rm AM}$ and $\vec v_{\rm AM}$ is evaluated exactly with
\begin{equation}
    \tau(\vec r_{\rm AM},\vec v_{\rm AM}) = \frac{1}{n_{\rm He}(\vec r_{\rm AM}) K(\vec r_{\rm AM},\vec v_{\rm AM})}\,,
    \label{exact_gamma}
\end{equation}
and
\begin{eqnarray}
  K(\vec r_{\rm AM},\vec v_{\rm AM}) & = &
   \int \mathrm{d}^3v_{\rm He}\, v_{\rm rel} \sigma_{\textrm{AM--He}}(E)
    \frac{P_{\rm He}(\vec r_{\rm AM},\vec v_{\rm He}) }{p_{\rm He}(\vec r_{\rm AM})}
    \nonumber\\
   &=& {\cal K}(|\vec v_{\rm AM}-v_{\rm He,tv}\hat r_{\rm AM}|, T_{\rm He}(r_{\rm AM}))
   \,,
    \label{v_weighted_sigma}
\end{eqnarray}
where
\begin{eqnarray}
   {\cal K}(v,T) &=& 4\pi \left(\frac{m_{\rm He}}{2\pi  k T}\right)^{3/2} e^{-m_{\rm He} v^2/(2kT)} \\
     && \times    \int_0^\infty v^2_{\rm rel} {\rm d} v_{\rm rel} 
         \left[ v_{\rm rel} \sigma_{\textrm{AM--He}}(E)\right] \nonumber \\
          && \qquad\qquad \times
          e^{-m_{\rm He} v_{\rm rel}^2/(2kT)}
           \frac{ \sinh[m_{\rm He} v_{\rm rel} v/kT]}{m_{\rm He} v_{\rm rel} v/kT}\,.
           \nonumber
\end{eqnarray}
Here, we  realize that $K(\vec r_{\rm AM},\vec v_{\rm AM})$ is  only a function of the local speed
of the alkali-metal atom in the frame moving along with velocity $v_{\rm He,tv}\hat r_{\rm AM}$
and the local helium temperature $T_{\rm He}(r_{\rm AM})$.
For efficient Monte-Carlo simulations, the rate coefficient ${\cal K}(v,T)$ is precomputed on a grid of $T$ from 0.1 mK to 100 mK  
and $v$ from 0 m/s to $v_{\rm cutoff}$. A two-dimensional interpolator is then used to determine ${\cal K}(v,T)$ at any $v$  and $T$  within the boundary of the grid. The grid size is chosen such that the difference between the interpolated  value and that from numerical integration of Eq.~\ref{v_weighted_sigma} is less than 1\,\%.
Moreover, the difference between Eqs.~(\ref{approx_gamma}) and (\ref{exact_gamma}) is less than 2\,\% for $v > v_{\rm cutoff}$ and $T_{\rm He} < 100$ mK, but Eq.~(\ref{approx_gamma}) is  an order of magnitude faster to compute. 
 
We use an acceptance-rejection procedure to determine whether an alkali-metal atom, located at phase-space point $(\vec r_{\rm AM}, \vec v_{\rm AM})$, collides with a $^4$He atom. 
This procedure starts by computing the local mean-free time $\tau_{\rm l}$ and equate time step $\Delta t$
to the preliminary value $ s\tau_{\rm l}$ with $0<s\ll1$. Our value for $s$ is discussed below.
The time step $\Delta t$ is adjusted if one of two conditions is met.
First,  if $v_{\rm AM}\Delta t > \Delta S$ for  distance $\Delta S$ discussed below, we set $\Delta t= \Delta S/v_{\rm AM}$.
We then compute $\tau_{\rm n}$, the local mean-free time at the ``next'' phase-space position $(\vec r_{\rm AM}+\vec v_{\rm AM}\Delta t, \vec v_{\rm AM})$ and if $\tau_{\rm l}\ge 2\tau_{\rm n}$, we half $\Delta t$ and the process of adjusting $\Delta t$ repeats.
The process  halts when neither inequality is met and we accept $\Delta t$.
We observe that $s\tau_l$ is the upper bound to the accepted time step.

The constraint $v_{\rm AM}\Delta t<\Delta S$ is required as alkali-metal atoms entering the jet far from the jet nozzle are in regions of relatively low $^4$He number density. Using only the local mean-free time to determine $\Delta t$ can incorrectly result in a particle traveling through the jet without undergoing collisions.

Finally, the probability of a collision between an alkali-metal atom and a $^4$He atom
is $p_{\rm AM}=(1/\tau_{\rm l}+1/\tau_{\rm n})\Delta t/2$. 
In our simulations, we use $s=0.1$ and $\Delta S= 1$ mm.
These values for $s$ and $\Delta S$ ensure that the likelihood of a collision is always less that 15\,\%  and
that we can compute sufficiently accurate statistical averages within a reasonable amount of time on the  computational resources available to us.

A random number $\cal P$ between 0 and 1 is now generated from the uniform probability distribution. If ${\cal P}\ge p_{\rm AM}$ then the
alkali-metal atom moves from $(\vec r_{\rm AM}, \vec v_{\rm AM})$ to $(\vec r_{\rm AM}+\vec v_{\rm AM}\Delta t, \vec v_{\rm AM})$.
If ${\cal P}< p_{\rm AM}$ a AM-$^4$He collision occurs at $(\vec r_{\rm AM}, \vec v_{\rm AM})$. We then generate a helium velocity $\vec v$ sampled from distribution $P_{\rm He}(\vec r_{\rm AM}, \vec v)$ in Eq.~(\ref{eq:PHelium})  for velocities $|\vec v_{\rm AM}-v_{\rm He,tv}\hat r_{\rm AM}|> v_{\rm cutoff}$, while for other velocities we sample from $v_{\rm rel}\sigma_{\rm AM-He}(E)P_{\rm He}(\vec r_{\rm AM},\vec v)$. This latter sampling technique is the DBRC algorithm described in Ref.~\cite{Romano_2018_DBRC} and ``favors'' $^4$He velocities where $v_{\rm rel}\sigma_{\rm AM-He}(E)$ is large.

Collisions change the velocities of the atoms conserving the center-of-mass velocity $\vec v_{\rm cm}$ and the length of the relative velocity $v_{\rm rel}=|\vec v_{\rm rel}|$, while keeping the atoms at the same location $\vec r_{\rm AM}$. The final velocities $\vec w_{\rm AM}$ and $\vec w$ for the alkali-metal atom and $^4$He, respectively, are most conveniently evaluated in  center-of-mass and relative coordinates with scattering angle $\theta_{\rm c}\in[0,\pi]$ sampled from $\sin\theta_{\rm c}{\rm d}\sigma_{\rm AM-He}/{\rm d}\Omega$ at the relative collision energy $E=\mu v_{\rm rel}^2/2$ and azimuthal angle $\varphi_{\rm c}$ uniformly sampled from $[0,2\pi]$. 
Some thought then shows that
the final velocity of the alkali-metal atom is $\vec w_{\rm AM}=\mu \vec w_{\rm rel} / m_{\rm AM}+ \vec v_{\rm cm}$, where the final relative velocity
\begin{equation}
  {\vec w}_{\rm rel}= (\sin\theta_{\rm c}\cos\varphi_{\rm c}  \hat x'
          +\sin\theta_{\rm c} \sin\varphi_{\rm c}  \hat y'
         + \cos\theta_{\rm c}  \hat z' ) v_{\rm rel}
\end{equation}
with  unit vector  $\hat z'=\hat v_{\rm rel}=(n_{{\rm rel},x},n_{{\rm rel},y},n_{{\rm rel},z})$ parallel to the initial relative velocity and 
unit vector $\hat x' = ( -n_{{\rm rel},y},n_{{\rm rel},x},0)/\sqrt{n^2_{{\rm rel},x}+n^2_{{\rm rel},y}}$
perpendicular to $\hat z'$.
Finally, unit vector $\hat y'=\hat z' \times \hat x'$ so that the three unit vectors form a positively oriented orthonormal basis.

The steps of the simulation are then repeated until the particle leaves our spatial bounds. The typical bounds of the simulation are $x\in[-6\,{\rm cm},6\,{\rm cm}]$, $y\in[-6\,{\rm cm},6\,{\rm cm}]$ and $z\in [-4\,{\rm cm},10\,{\rm cm}]$.
These bounds are motivated by the $^4$He jet density profile and the locations of the atomic oven and skimmer. The net result is an ensemble of computed trajectories from which observable quantities can be calculated. The number of computed trajectories $N$  varies from a few million to 100 million for converged results.
A larger $N$ is required for smaller spatial regions or velocity intervals. The simulation is written in Python and utilizes the Numba library for optimized performance \cite{numba}. 

\section{Results}

\subsection{Seeding with \texorpdfstring{$^7$}{7}Li}

\begin{figure}
    \includegraphics[width=1.0\columnwidth]{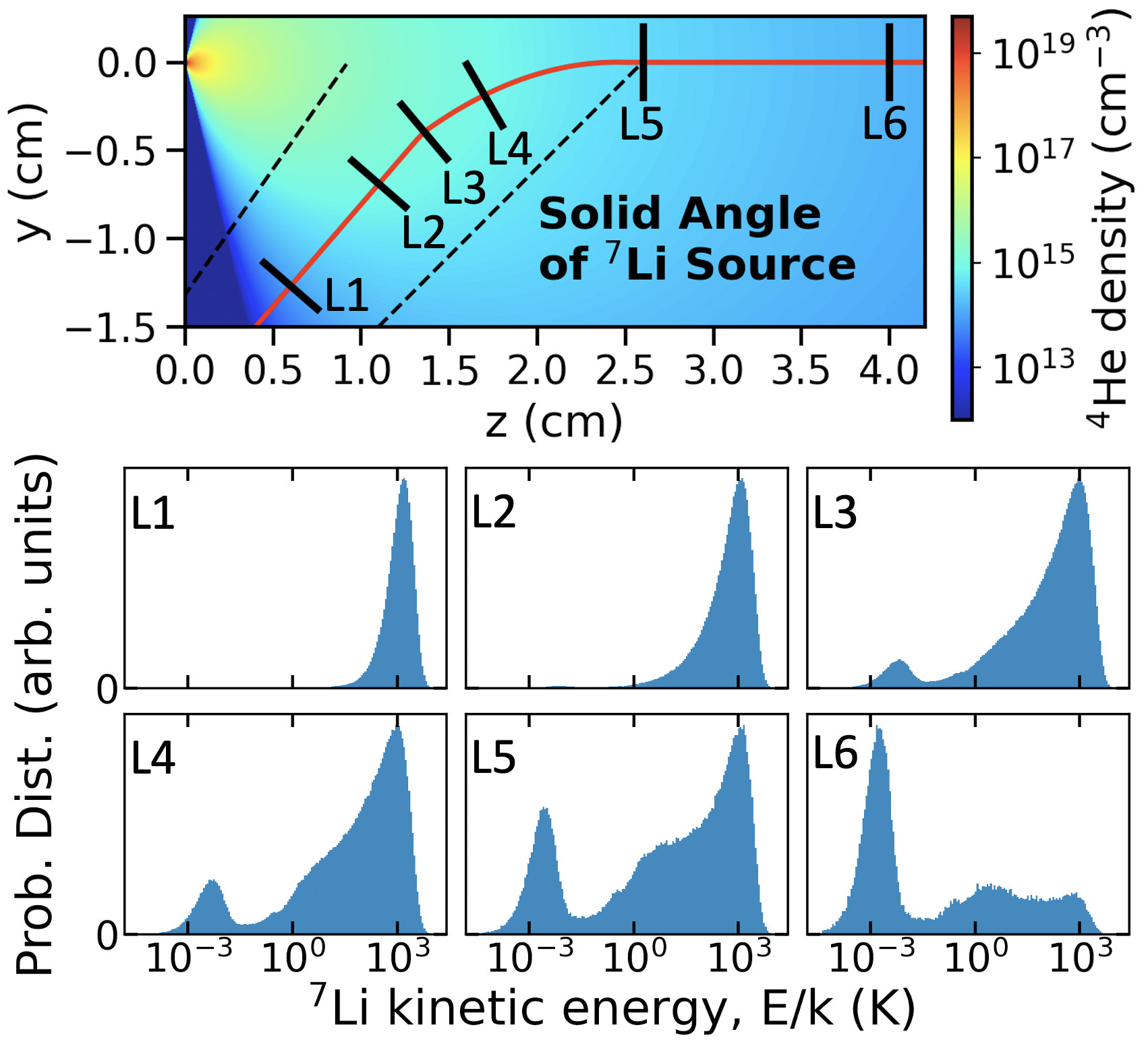}
    \caption{(Top panel) Representative trajectory of seeded $^7$Li atoms (red curve) from our 3D Monte-Carlo simulations overlaid on a 2D cut through the number density profile of the $^4$He jet at 200 sccm for the $(0,y,z)$ plane. The sonic nozzle of the $^4$He jet is located at $(y,z)=(0,0)$. The trajectory has been post-selected to lie within planes $x=-0.1$ cm and 
    $x=0.1$ cm.  The dashed lines indicate the solid angle within which most $^7$Li atoms from the atomic oven enter our simulation region. 
    (Bottom panels labelled L$j$ with $j=1$, 2, 3, 4, 5, and 6)
    $^7$Li number probability distributions as functions of the $^7$Li kinetic energy in a  moving frame of the $^4$He jet  for various small areas L$j$ along but perpendicular to the trajectory shown in the top panel. The vertical axes of the six panels are on different scales and should not be compared.
    }
    \label{Li_cooling}
\end{figure}

The representative trajectory of seeded $^7$Li atoms which exit the skimmer projected onto the $yz$ plane is shown in the top panel of Fig.~\ref{Li_cooling} for a $^4$He flow rate of 200 sccm. Here,  a $^4$He flow rate of 1 sccm, an abbreviation for standard cubic centimeters per minute,  corresponds to $\dot{N}_0=4.48\times 10^{17}$ atoms/s.} (The term standard in unit sccm reflects standard conditions for temperature and pressure of 273.15 K and 101.325 kPa, respectively.)
Places along the trajectory with a larger curvature imply places with a higher rate of collisions with helium atoms.
The six panels at the bottom of Fig.~\ref{Li_cooling} show distributions of the kinetic energy $E=m_{\rm Li} (\vec v_{\rm Li}-v_{\rm He,tv}\hat r)^2/2$ of $^7$Li atoms in the frame moving along with the local terminal velocity of  helium atoms in the jet. Each panel shows a distribution of kinetic energies of atoms as they pass through 
a rectangular region perpendicular to and centered on points along the trajectory in the top panel of Fig.~\ref{Li_cooling}. The region extends by $\pm 0.1$ cm into the $x$ direction. Its length in the $y$-$z$ plane is 0.4 cm.

\begin{figure}
   \includegraphics[width=\columnwidth]{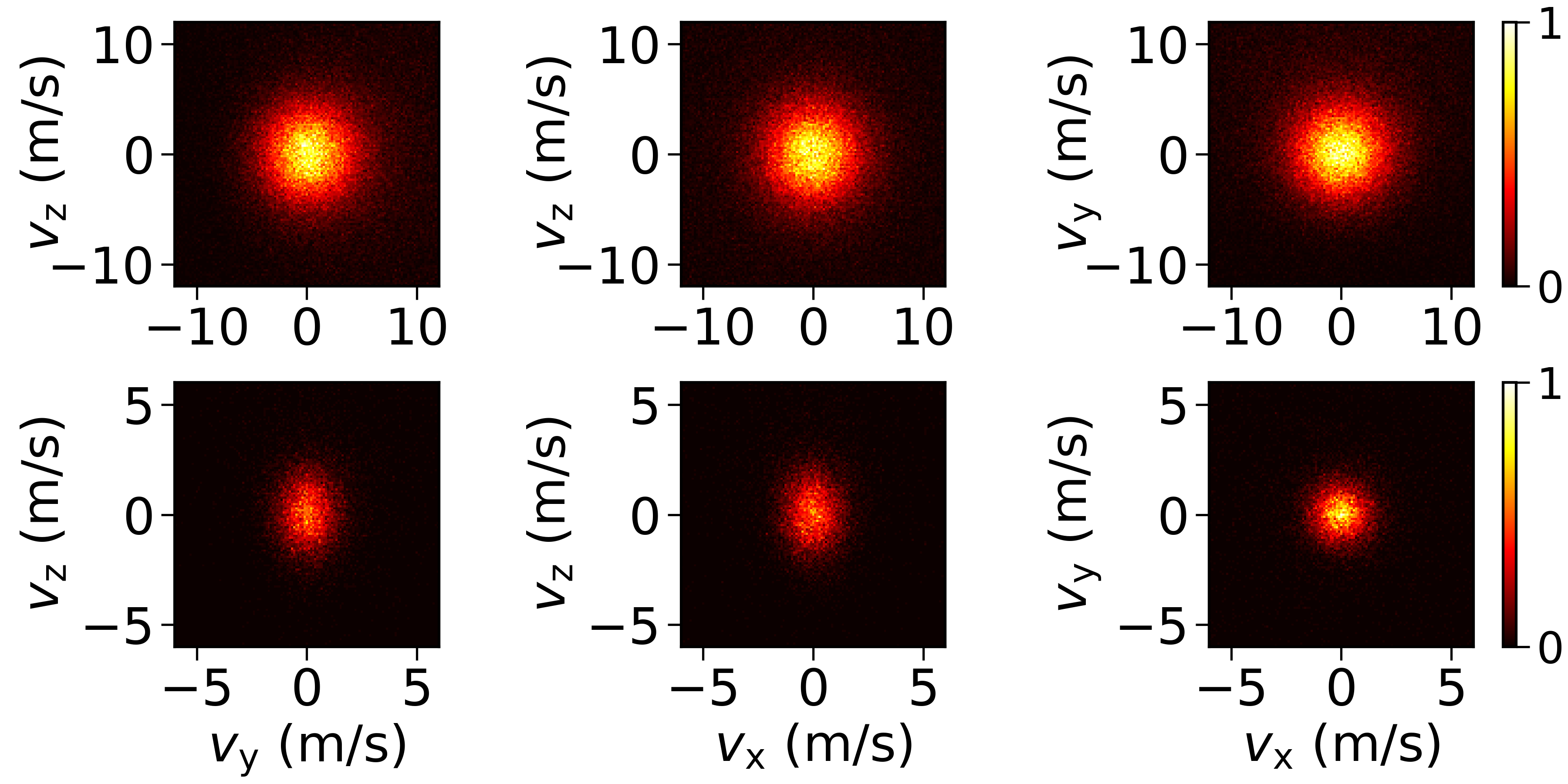}
    \includegraphics[width=\columnwidth]{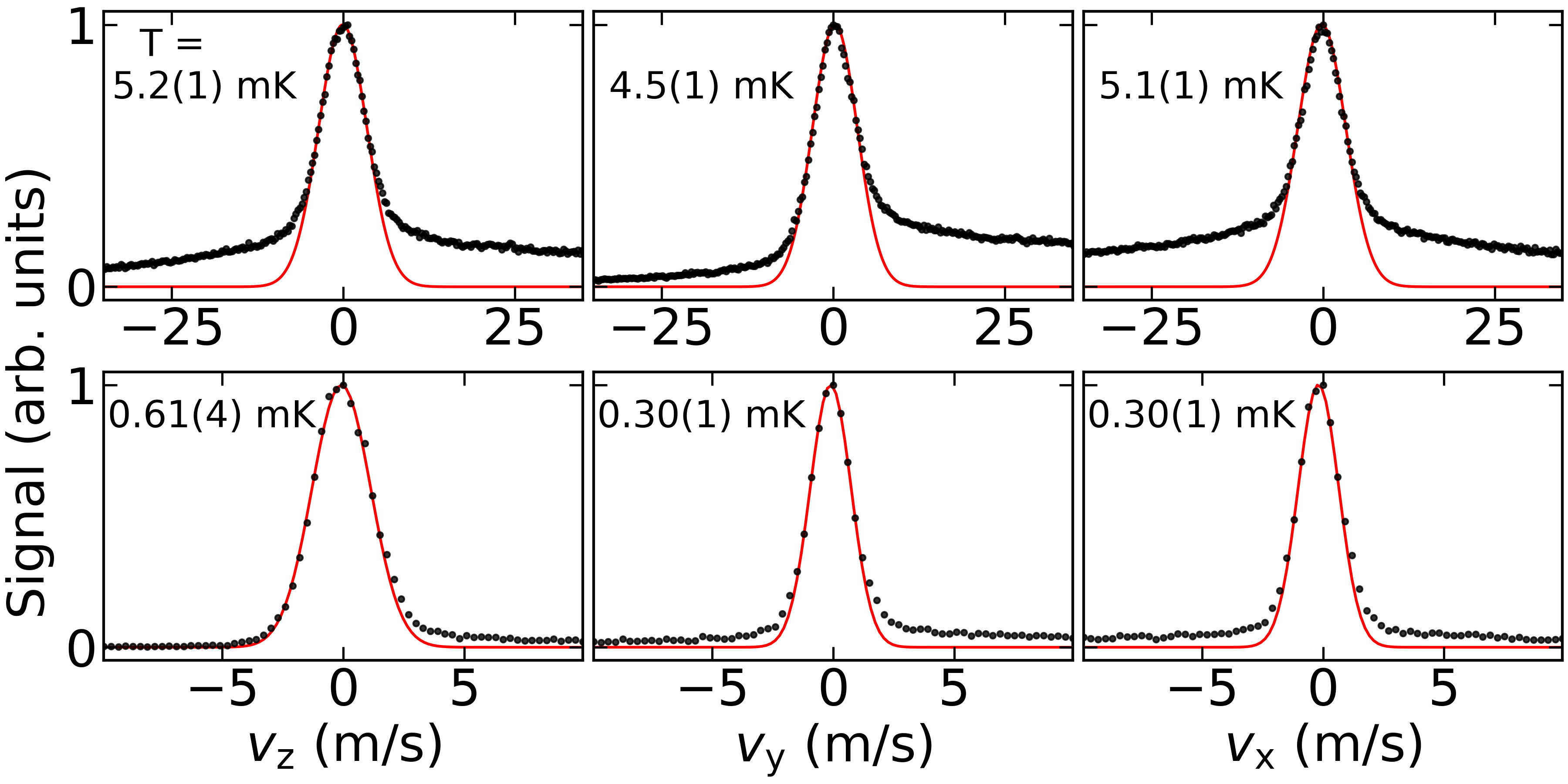}
    \caption{
    Contour and line graphs of 2D and 1D cuts through  3D $^7$Li velocity distributions 
for two positions along the center line of the jet at a  helium flow rate of 200 sccm.
The velocity distributions are relative to the terminal velocity of the $^4$He atoms at these positions.
We have combined data from all trajectories passing within a 1 mm radius circle perpendicular to the center line.  The top row is for a distance of 2 cm from the sonic nozzle while the bottom row is for a distance of 10 cm.
 Simulated $^7$Li velocity distributions (black dots) and Gaussian profiles fit to the FWHM of the central peak (red curves) as functions of velocity components $v_z$, $v_y$, and $v_x$ for the left, middle, and right columns, respectively.
    The top row of panels shows velocity distributions taken along the center line of the helium jet at a distance of 2 cm from the nozzle, while the bottom row is at a distance of 10 cm. For each line graph, we specify the fitted FWHM velocity $v_{\rm fit}$ as $T_{\rm fit}=m_{\rm Li} v^2_{\rm fit}/(2k)$.}
    \label{Li:vel_2cm_10cm}
\end{figure}

The distribution of kinetic energies as the $^7$Li atoms pass through area L1 and begin to enter the jet is a Maxwell-Boltzmann distribution with a temperature of $800$ K, equal to the effusive source temperature.
For L2, the $^7$Li atoms have started to cool as the distribution becomes more pronounced for smaller $E$.
At the third region, a small second peak at $E/k\approx 10$ mK has appeared indicating that $^7$Li atoms are becoming trapped in the jet. This second peak becomes more pronounced for the last three positions along the trajectory with its peak position shifting to a smaller $E$. The captured $^7$Li atoms are getting colder.
It should be recognized that as the L$j$ have a finite area, the increasing prominence of the peak 
near $E/k\approx 1$ mK relative to those at higher kinetic energies cannot used to estimate the jet capture efficiency. For example, for L6, an area nearly perpendicular to the center line of the jet, we do not account for many of the  $^7$Li atoms that fail to be captured. Those atoms will  have a large  kinetic energy. 

\begin{figure}[t]
    \centering
 
    \includegraphics[width=0.95\columnwidth]{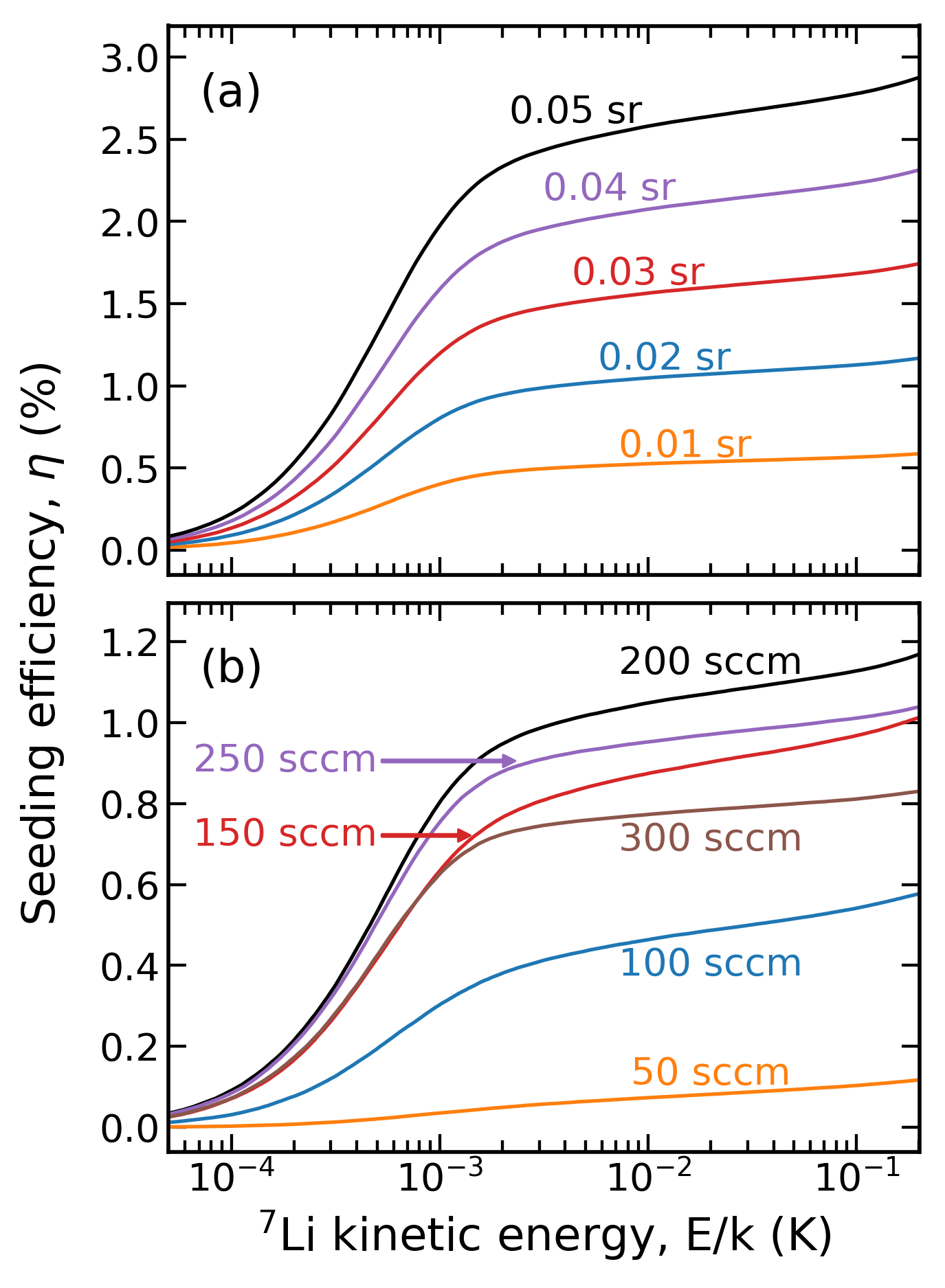}
    \caption{Percentage of simulated $^7$Li atoms with energy in the moving frame up to the value specified on the x-axis. Results are at a distance of 10 cm from the sonic nozzle with the solid angle defined relative to the location of the sonic nozzle. (a) Results for various solid angles with a helium flow rate of 200 sccm. (b) Results for various helium flow rates and a solid angle of 0.02 sr.
    }
    \label{seeding_efficiency_Li}  
\end{figure}

Figure~\ref{Li:vel_2cm_10cm} shows several cuts through simulated three-dimensional $^7$Li velocity distributions  2 cm and 10 cm away from the sonic nozzle along the center line of the jet and with a $^4$He flow rate of 200 sccm. First, not unexpectedly given the complex  device geometry in Fig.~\ref{apparatus}, the velocity distributions are asymmetric, can have large wings, and are different along the three directions. 
Second, the lithium atoms velocity distributions become narrower for larger distances away from the sonic nozzle and, finally, the central peak is more prominent relative to the wings.

As the $^7$Li kinetic energy and velocity profiles do not follow a Maxwell-Boltzmann distribution, extracting a temperature is dubious. Nevertheless, we can define an effective temperature for the captured atoms by fitting the central peaks to Gaussian distributions as shown in Fig.~\ref{Li:vel_2cm_10cm}. At a distance of 10 cm from the nozzle, results indicate similar temperatures as that of the helium. It is desirable to know what percentage of the injected $^7$Li atoms cool to these effective temperatures. As this is complicated by the asymmetry of the velocity distributions we instead quantify the cumulative fraction of total simulated particles with a kinetic energy in the moving frame less than a specified value $E$. We define the seeding efficiency $\eta(E, \Omega)$ as this fraction of particles which are also directed within a solid angle $\Omega = \pi r^2/L^2$; i.e. through a circle of radius $r$ centered on the beam axis at a distance $L$ from the nozzle.

In Fig.~\ref{seeding_efficiency_Li}, the seeding efficiency is given for a variety of flow rates and solid angles at a distance of $L=10$ cm from the nozzle. A distance of 10 cm is chosen as most of the cooling has occurred by this point. Additional discussion of cooling versus distance from the nozzle can be found in the appendix. For our source geometry, the skimmer can extract atoms within a solid angle of $\approx 0.02$ sr. As simulations show that the seeding efficiency increases with solid angle, a larger extractable flux could be obtained with a different skimmer geometry. However, this also results in a larger flux of helium atoms leaving the skimmer. For reference, with a solid angle of $\approx 0.02$ sr, a helium flow rate of 200 sccm, and the density distribution given by Eq.~\ref{eq:HeDensity}, the gas load of helium leaving the skimmer is $\approx 0.033$ Torr L/s. In practice, careful consideration should be taken to ensure acceptable vacuum pressures outside of the cryogenic region.

\subsection{Comparison with \texorpdfstring{$^7$}{7}Li experiments}

\begin{figure*}
    \includegraphics[width=2\columnwidth]{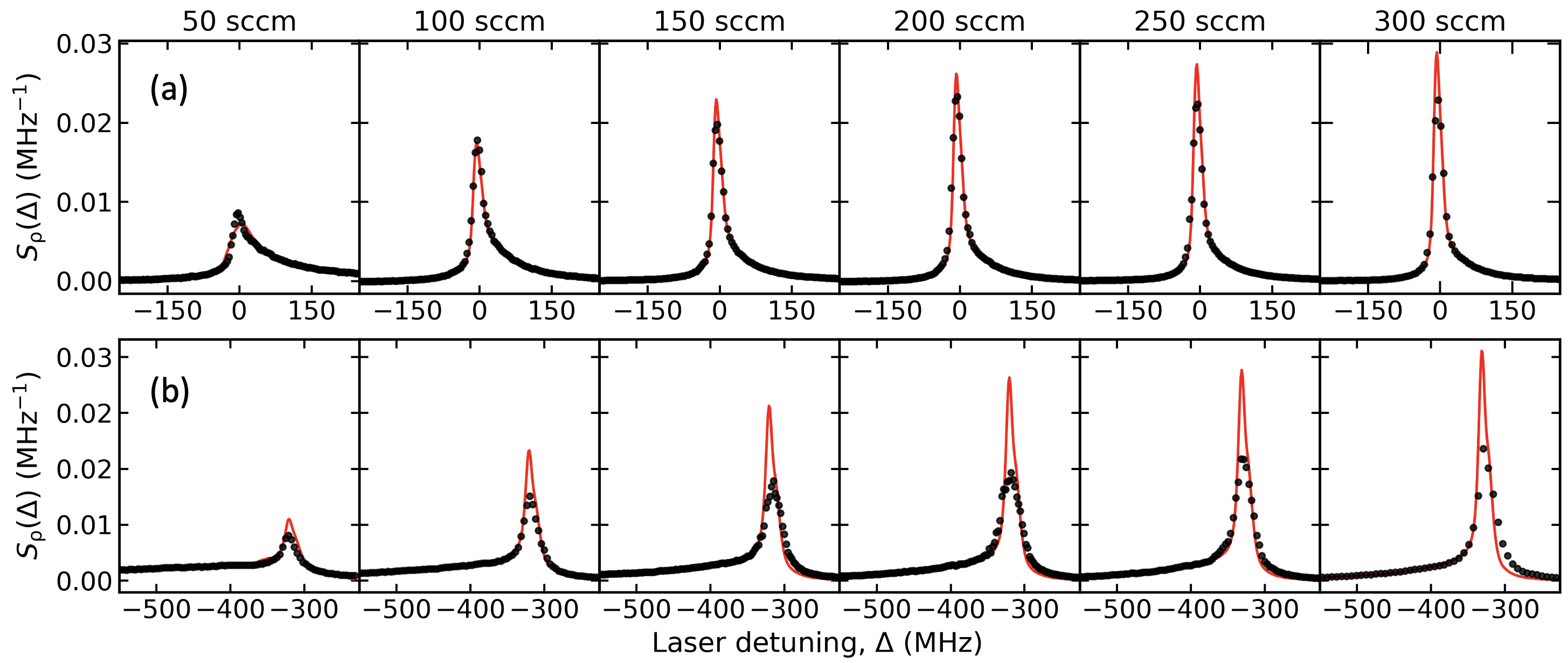}
    \caption{Experimental $^7$Li fluorescence spectra (black markers) and  simulated spectra (red curves) as functions of 
    the laser detuning $\Delta$ from the average transition frequency of the $F=2$ to $F'=1,2,3$ transitions of the D$_2$ line of $^7$Li weighted by the respective transition strength. 
     Transverse spectra taken $4.1$ cm downstream from the nozzle are shown along the top row (a), while longitudinal spectra are shown along the bottom row (b). From left to right spectra are taken for six, increasing $^4$He flow rates,  indicated above the top row in units of sccm. }
    \label{transerveProfiles}
\end{figure*}

In the experimental realization, studied in Ref.~\cite{Huntington_2022_Cold_Atom_Source}, local $^7$Li density and velocity distributions were determined with fluorescence spectroscopy on transitions between magnetic sublevels of the $F=2$ hyperfine state of the electronic 1s$^2$2s($^2$S) ground state to  three of the four hyperfine components $F'$ of the excited  1s$^2$2p($^2$P$_{3/2}$) state. This is the D$_2$ line in $^7$Li. We measure laser-induced fluorescence as a function of laser frequency $\nu$ either with a linearly polarized probe laser parallel to the center line of the jet and propagation direction $-\hat z$  (laser beam $\rho = \parallel$) or a linearly polarized probe laser perpendicular to the center line and propagation direction $+\hat y$ (laser beam $\rho = \perp$). The  laser perpendicular to the jet is located 4.1 cm from the sonic nozzle. 

In our simulations, the laser spatial intensity profile $I_\rho(\vec r)$ of  lasers $\rho=||$ and $\perp$, propagating parallel and perpendicular to the center line of the jet, respectively, are modeled as rectangles with a width along $x$ equal to twice the beam waist, $w_0=0.41$ cm, along their propagation direction. The height of the rectangle along $y$ is generally chosen to be small such that variations in the laser intensity can be neglected. Inside the rectangle, the intensity is constant while outside it is zero. Moreover, the lasers operate at sufficiently low intensities that optical pumping effects are negligible.

From the ensemble of trajectories, we compute the distribution $P_{\rm AM}(x,y,\vec{v}_{\rm AM})$ of positions $x$ and $y$ at which particles pass through a surface perpendicular to the jet axis at a distance $L$ from the nozzle, with $\vec{v}_{\rm AM}$ their velocity at that position. 

The simulated spectral profile for laser beam $\rho$ is
\begin{eqnarray}
   \lefteqn{ S_\rho(\Delta)= \kappa \int_{A} {\rm d}x{\rm d} y\int {\rm d}^3\vec v_{\rm AM}   P_{\rm AM}(x,y, \vec v_{\rm AM}) } 
   \label{eq:lineshape}\\
    && \times  W_\rho(\vec v_{\rm AM}) 
        \sum_{F'=1}^3  \frac{ D_{F'} \ I_\rho(\vec r_{\rm AM}) }{1+4[\Delta-\vec k_\rho\cdot \vec v_{\rm AM}/h-\Delta_{\rm hf}(F')]^2/\gamma^2} \,,
        \nonumber
\end{eqnarray}

where $\kappa$ is a constant of proportionality. Since we do not simulate absolute signal levels, we adjust $\kappa$ such that $\int S_{\rho}\left(\Delta \right) \rm{d}\Delta =1$ for both experiment and simulation.
$\Delta=\nu-E_{D2}/h$ is the laser detuning from the average transition frequency from $F=2$ to $F'=1,2,3$ weighted by the respective transition strength. The interrogation area $A$ is a rectangle with a height equal to $0.60(2)$ mm. The width is equal to the laser beam waist. We have found that results are insensitive to variations in the width of the rectangle indicating negligible variations in the velocity profiles $v_{\rm y}$ and $v_{\rm z}$ along $x$. As such we do not account for the laser intensity profile along $x$ for computing simulated spectral profiles. Here, $h$ is the  Planck constant.
Moreover, the sum is over the optically allowed hyperfine levels $F'$ of the 1s$^2$2p($^2$P$_{3/2}$) state, and $\Delta_{\rm hf}(F')=+9.619$ MHz, $+3.730$ MHz, and $-5.656$ MHz are hyperfine frequency shifts for the $F'=1$, 2, and 3 levels of the $^7$Li 1s$^2$2p($^2$P$_{3/2}$) state, respectively \cite{Das_2007_D2_Lines}.
The relative strengths $D_{F'}$ are $1/20$, $1/4$, and $7/10$ for $F' = 1$, 2 and 3, respectively, to account for the different line strengths of the $F=2$ to $F'$ transitions. The natural linewidth of the $^7$Li 1s$^2$2p($^2$P$_{3/2}$) state is $\gamma=5.87$ MHz \cite{Li_NaturalLinewdith}.

The inner product $\vec k_\rho\cdot \vec v_{\rm AM}$ in the denominator of the Lorentzians in Eq.~(\ref{eq:lineshape}) accounts for  Doppler shifts. Here, unit vector $\hat  k_\rho=\vec k_\rho/k_\rho$ gives the direction of  laser $\rho$ and wavenumber $k_\rho$ is given by the dispersion relation $h\nu=\hbar c k_\rho$ with  speed of light in vacuum $c$  and  reduced Planck constant $\hbar$. It is worth remembering that the temperature of  captured alkali-metal atoms is much
smaller than $m_{\rm AM} v_{\rm He,tv}^2/(2k)$ and for laser beam $\rho=||$ and $^7$Li atoms, $k_{||}v_{\rm He,tv}/h=310$ MHz.
Weight function $W_\rho(\vec v_{\rm AM})$ in Eq.~(\ref{eq:lineshape}) accounts for the reduced number of fluorescence photons emitted from faster-moving atoms that spend less time in the laser beam. 
For both parallel and perpendicular lasers, we use $W_\rho(\vec v_{\rm AM})=1/v_{\rm AM}$ as the detected volume 4.1 cm from the sonic nozzle is small.

Line-shapes for both experiment and simulation are given in Fig. \ref{transerveProfiles}. Line-shapes at this distance show a clear asymmetry. Simulated profiles reproduce the measured asymmetry while also predicting a narrower central peak. For the longitudinal profiles, we estimate an optical density of $\approx$0.3 over the distance of our chamber ($\approx$2 m). The non-negligible absorption effects may in part explain the larger discrepancy for the longitudinal profiles. 

One additional source of broadening that we have not included is laser frequency noise. The FWHM of the laser line-width over the integration time of the atoms is about 2 MHz. As this is small compared to the other broadening mechanisms, there is no apparent change in the simulated spectral profiles when $S_{\rm \rho}(\Delta)$ is convolved with the laser noise profile.

A point of interest in our apparatus is maximizing seeded lithium that is extracted by a 2.54 cm diameter skimmer located 16 cm from the nozzle as shown in Fig. \ref{apparatus}. Simulations indicate that flow rates around 200 sccm for a seeding distance of 1.7 cm maximize extractable lithium. To compare to experiment, the injected lithium flux is measured with zero helium flow. This allows us to compute the average simulated alkali-metal atom number density that is within the projected area $A$ of the skimmer using the following:
\begin{equation}
    n_{\rm sim} = \frac{\Phi}{NA} \sum_i^{N_{c}} \frac{1}{v_i}\,,
    \label{densityEq}
\end{equation}
where $\Phi$ is the measured experimental alkali-metal-atom flux entering the cryogenic region, $N$ is the total number of simulated alkali-metal particles, $v_i$ is the speed of simulated particles within the capture region, and $N_{c}$ is the number of simulated particles within $A$. We compare the simulated and measured total density at all velocities. This comparison is made because the experimental measurements cannot distinguish if an atom is within a narrow velocity range in both the transverse and longitudinal directions. 

\begin{figure}
\includegraphics[width= 0.9\columnwidth]{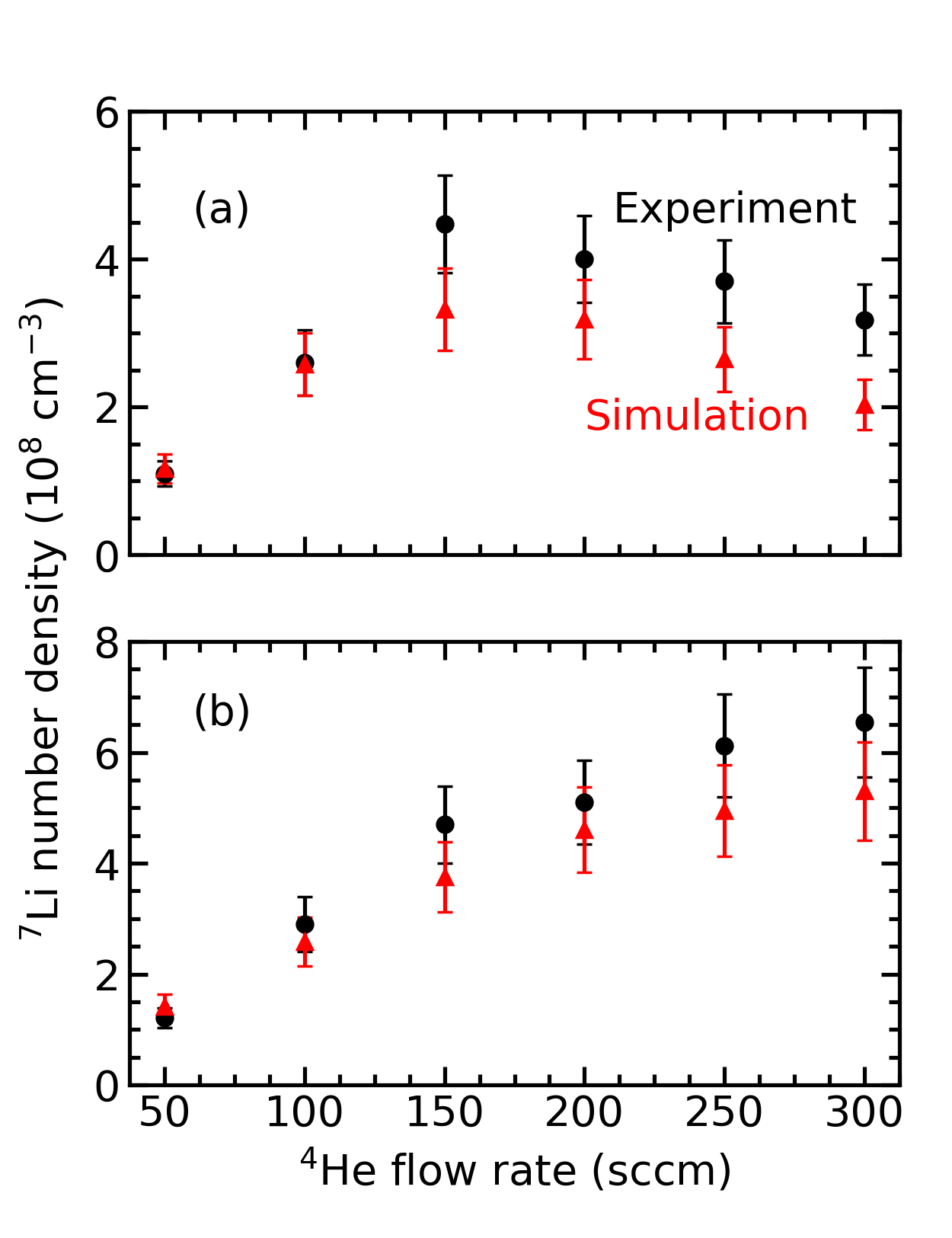}
   \caption{Measured (black circles) and modeled (red triangles) average $^7$Li number densities $n_{\rm Li}$ at 4.1 cm from the nozzle within a circle of radius $\approx 0.33$ cm as a function of helium flow rate. Results in (a) correspond to a circle centered at $(x,y) = (0,0)$ while (b) corresponds to a circle centered at the location of the peak number density along $y$ with $x=0$. The terminal velocity of the lithium beam is 210(2) m/s for a helium flow rate below 220 sccm and 217(2) above 220 sccm. The simulated $^7$Li number densities follow from the measured flux leaving the lithium source of $\SI{1.4(2)e14}{}~\mathrm{s}^{-1}$ at a lithium oven temperature of 800 K.}
    \label{density_measurements}
\end{figure}

Results using our transverse probe for a variety of helium flow rates at a distance of 4.1 cm from the nozzle are given in Fig.~\ref{density_measurements}. At this distance, the projection of the skimmer corresponds to a circle of radius $\approx 0.33$ cm. In Fig.~\ref{density_measurements}(a), the number density for a circle centered at $(x,y) = (0,0)$ is given with the simulation correctly predicting a maximum average density at a helium flow rate of $150$ sccm. This result may appear to contradict the seeding efficiencies given in Fig.~\ref{seeding_efficiency_Li}, however, it is important to note that the simulated and experimentally measured density includes all velocity classes extending well past the values shown in Fig.~\ref{seeding_efficiency_Li}. While a flow rate of $150$ sccm resulted in a larger density within $A$, the simulation predicts that a flow rate of $200$ sccm results in more seeded atoms with lower energies in the moving frame. 

\begin{figure}
\includegraphics[width=\columnwidth]{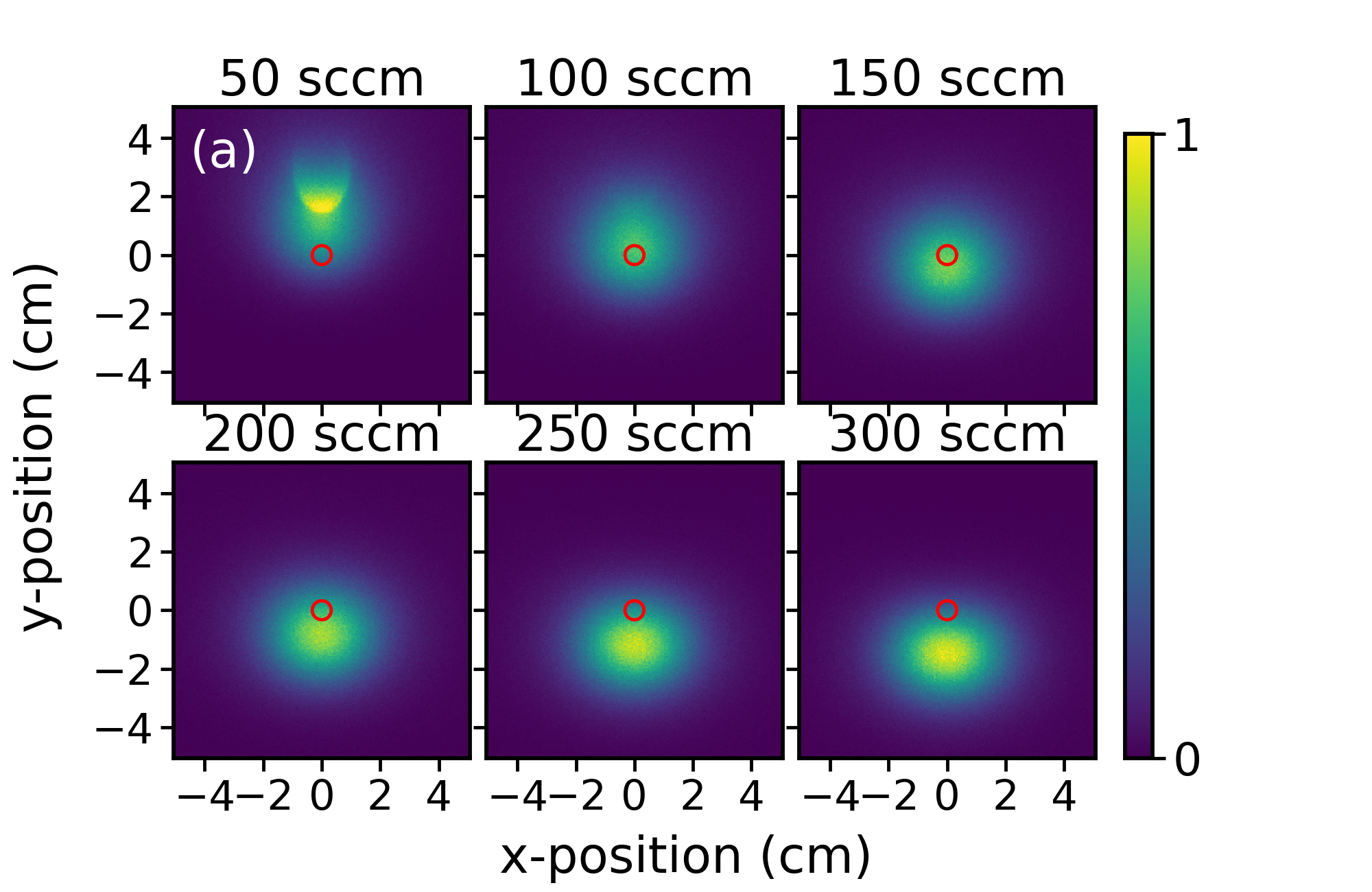}
\includegraphics[width=\columnwidth]{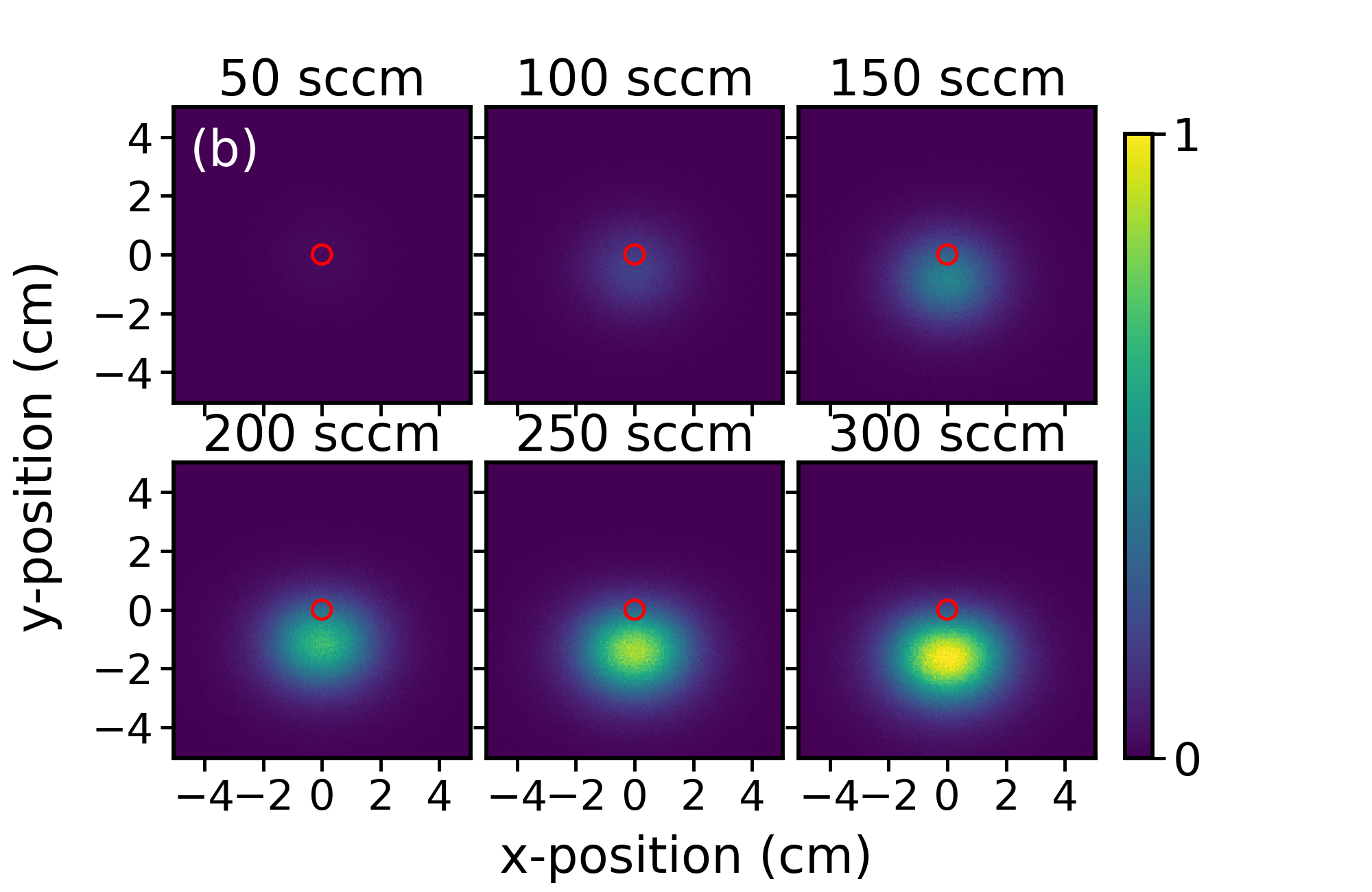}
   \caption{2D histogram of the positions of simulated $^7$Li atoms that pass the xy plane $4.1$ cm from the nozzle. The red circle indicates particles that are within the projection of the skimmer. The numbers above each plot indicates the $^4$He flow rate. (a) Results for all $^7$Li energies in the moving frame. (b) Results for $^7$Li energies $\le k\times$ 50 mK in the moving frame.}
    \label{simulationParticles}   
\end{figure}

Shown in Fig.~\ref{density_measurements} (b) is the number density but with the circle centered at the location of the peak density along $y$ with $x=0$. For flow rates below $\approx150$ sccm the location of the peak number density occurs above the center-line of the jet while for flow rates above $\approx150$ sccm it occurs below the center-line. Since the helium density along $y$ is largest at $y=0$, it may initially seem surprising that the location of the peak $^7$Li density is not necessarily located at $y=0$. This effect of helium flow rate on the position of the extractable lithium atoms is best visualized by examining the position of simulated particles. The positions of all particles which pass the xy plane $4.1$ cm in front of the nozzle are given by Fig.~\ref{simulationParticles} (a). At the highest flow rates particles are being deflected by the jet while at the lowest flow rate, the shadow of particles exiting the aperture in front of the oven is visible. In Fig.~\ref{simulationParticles} (b), we show the position of only particles with energies $\le k \times 50$ mK in the moving frame. Notably, if one wishes to maximize the extraction of thermalized lithium, it would be advantageous to not extract at the center-line of the jet but beneath it. 

For all simulated density results, we find that the uncertainties of the capture rates obtained with our MC results are larger than those of the differential cross-sections. This simply reflects the small theoretical uncertainties in the differential cross section for $E>10 $ K.

\begin{figure}[t]
    \centering
 
    \includegraphics[width= \columnwidth]{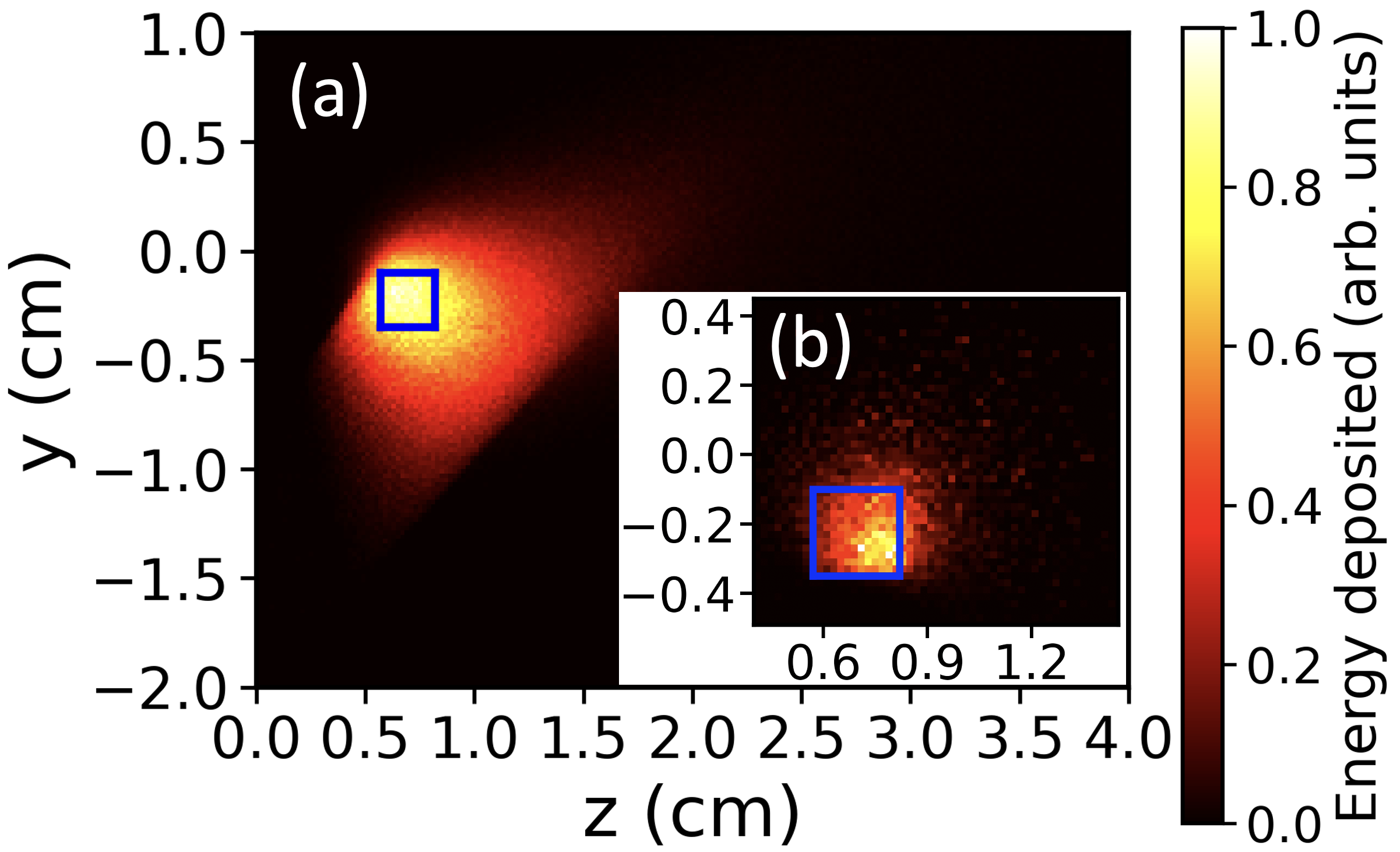}

    \caption{(a). Location and relative magnitude of energy deposited by $^7$Li atoms from $x=-0.1$ cm to $x=0.1$ cm. (b) Relative magnitude of energy deposited for the next He-He collision for helium atoms which underwent a collision with a lithium atom in the blue square.}
    \label{energyMap}  
    
\end{figure}

Due to the large reservoir of helium, we assume no appreciable heating of the jet and expect this to hold true for sufficiently low lithium flux. For reference, with a seeded flux on the order of $10^{14}$ s$^{-1}$, there are approximately 500,000 helium atoms per lithium atom for $150$ sccm of helium flow. For an oven temperature of $800$ K, the average energy of a lithium atom entering the jet is $\approx k\times1070$ K. If this energy were uniformly distributed among the helium atoms it would deposit only $k\times2$ mK per helium atom. To further examine how energy is deposited in the jet we record the location as well as the pre and post-collision velocity of the lithium and helium. A map of the energy deposition can then be constructed and is given by Fig \ref{energyMap} (a), which shows that most of the energy deposited by the lithium is deposited before the lithium reaches the center of the jet.

Naively it might seem like a large fraction of energy that is deposited in the periphery of the jet would remain there as the jet expands. This would limit heating effects along the center of the expansion. However, we have examined this further and found that it is only partially true. Depending on the energy transferred from the lithium atom and scattering angles, an energetic helium atom can have a significantly larger mean free path relative to the local mean free path of less energetic helium atoms. 

To quantify this effect, the simulation is repeated using the location and post-collision velocity of the helium within the blue square in Fig \ref{energyMap} using the $^4$He-$^4$He cross-section from Ref \cite{Chrysos_2017_He_He_cross_section}. The helium atoms then travel until they leave the simulation bounds or undergo a single collision. Fig. \ref{energyMap} (b) shows the energy deposition for this group of helium atoms. Helium atoms containing approximately $30\%$ of the energy deposited in the blue square leave the simulation bounds without undergoing further collisions. While this method does not allow us to determine how the remainder of the energy is ultimately distributed in the jet, it illustrates that a non-negligible amount of energy is effectively removed. It also suggests that heating of the center of the jet can come from energy that is initially deposited in the periphery. For sufficiently high lithium flux there will be heating of the jet and the assumptions made for the jet density and velocity distributions will no longer be valid. To date, we have been able to seed the jet with a lithium flux on the order of $10^{15}$ s$^{-1}$ and extract a lithium beam with a longitudinal temperature of $7(3)$ mK and a transverse temperature $< 20$ mK \cite{Huntington_2022_Cold_Atom_Source}. 

\subsection{Seeding with \texorpdfstring{$^{87}$}{87}Rb}\label{sec:Rb}

\begin{figure}
    \centering
    \includegraphics[width= 0.8\columnwidth]{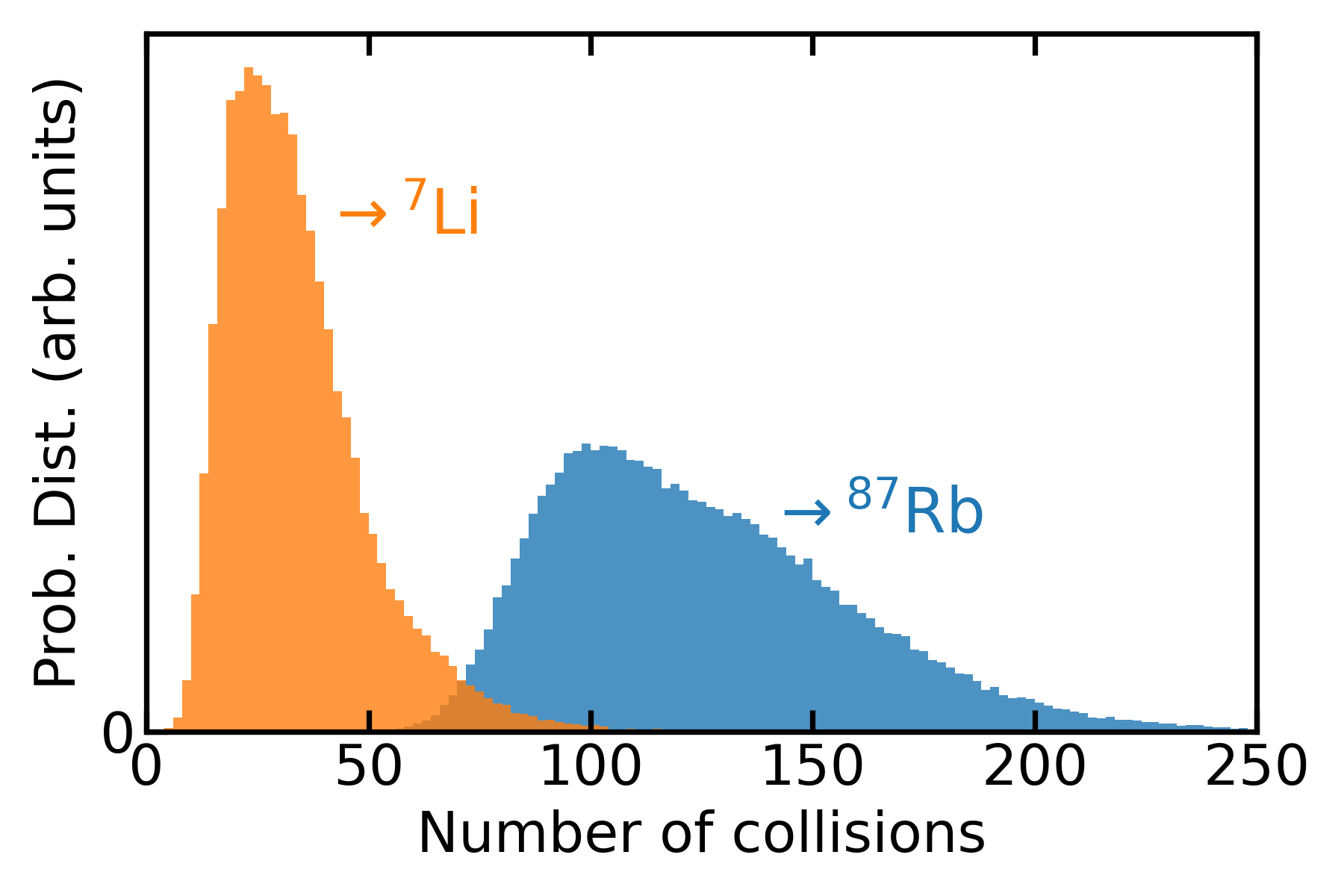}
    \caption{Probability distributions for the total number of $^4$He-$^7$Li and $^4$He-$^{87}$Rb collisions for atoms at a distance of 10 cm from the nozzle and within a solid angle of 0.02 sr. The $^4$He flow rate is 200 sccm for $^7$Li and 300 sccm for $^{87}$Rb.}
    \label{collision_comparison}  
\end{figure}

\begin{figure}
    \includegraphics[width=1.0\columnwidth]{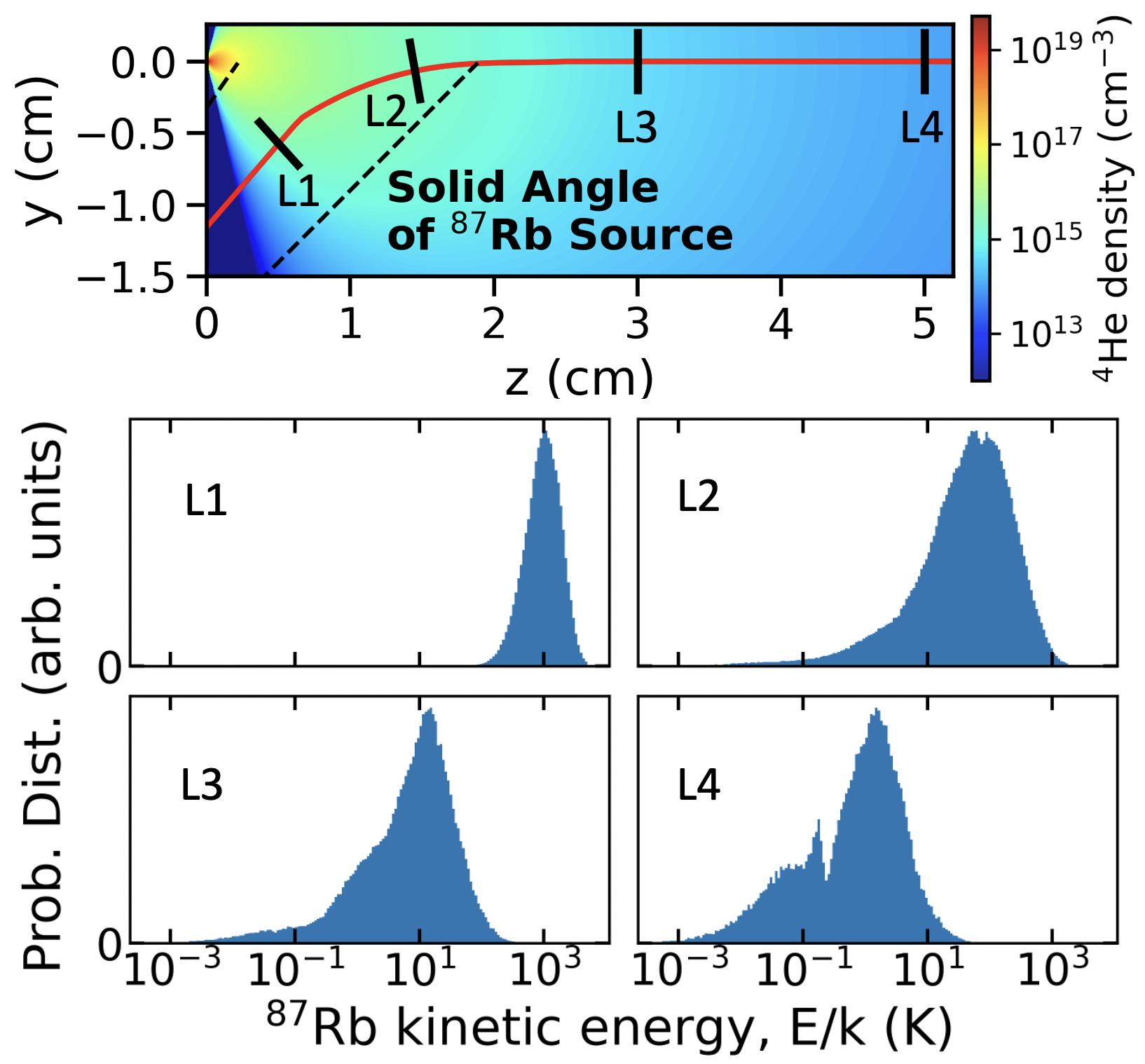}
    \caption{(Top panel)
    Representative trajectory of seeded $^{87}$Rb atoms (red curve) from our 3D Monte-Carlo simulations overlaid on a 2D cut through the number density profile of the $^4$He jet at 300 sccm for the $(0,y,z)$ plane. The sonic nozzle of the $^4$He jet is located at $(y,z)=(0,0)$. The trajectory has been post-selected to lie within planes $x=-0.1$ cm and 
    $x=0.1$ cm.  The dashed lines indicate the solid angle within which most $^{87}$Rb atoms from the atomic oven enter our simulation region. 
    (Bottom panels labelled L$j$ with $j=1$, 2, 3, 4)
    $^{87}$Rb number probability distributions as functions of the $^{87}$Rb kinetic energy in a  moving frame of the $^4$He jet  for various small areas L$j$ along but perpendicular to the trajectory shown in the top panel. The vertical axes of the four panels are on different scales and should not be compared.
    }
    \label{Rb_cooling}
\end{figure}

As our method is intended to be applicable for a variety of species, there is interest in examining the efficiency for seeding heavier particles into the jet. Here we present simulations of seeding $^{87}$Rb into the helium jet. We assume the same solid angle as our lithium source but at an oven temperature of 550 K due to rubidium's significantly higher vapor pressure. Due to having a higher initial momentum, seeding becomes more challenging as more collisions are required to thermalize the species into the jet. 

\begin{figure}
    \centering
    \includegraphics[width= \columnwidth]{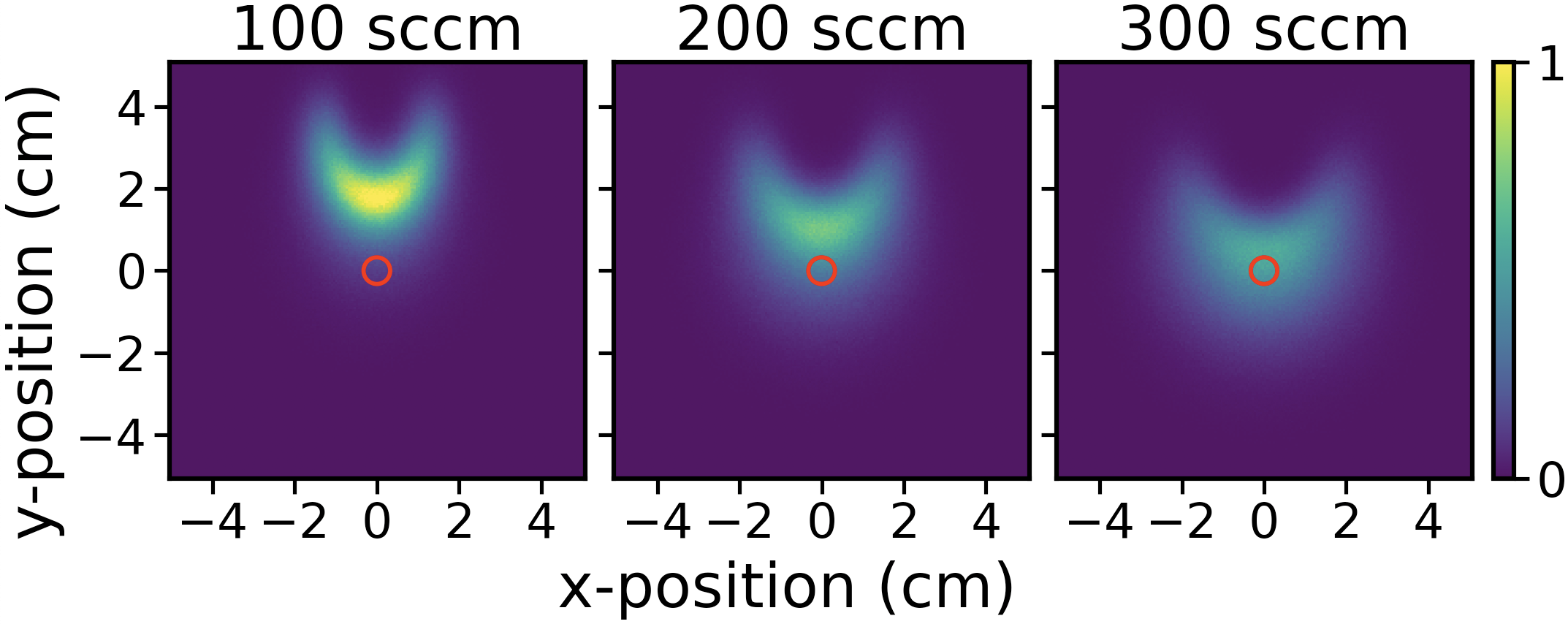}
    \caption{2D histogram of the position of simulated $^{87}$Rb atoms that pass the xy plane, 4.1 cm from the nozzle. The red circle indicates particles that are within the projection of the skimmer at this distance.}
    \label{simulated_particles_Rb}  
\end{figure}

\begin{figure}
   \includegraphics[width=\columnwidth]{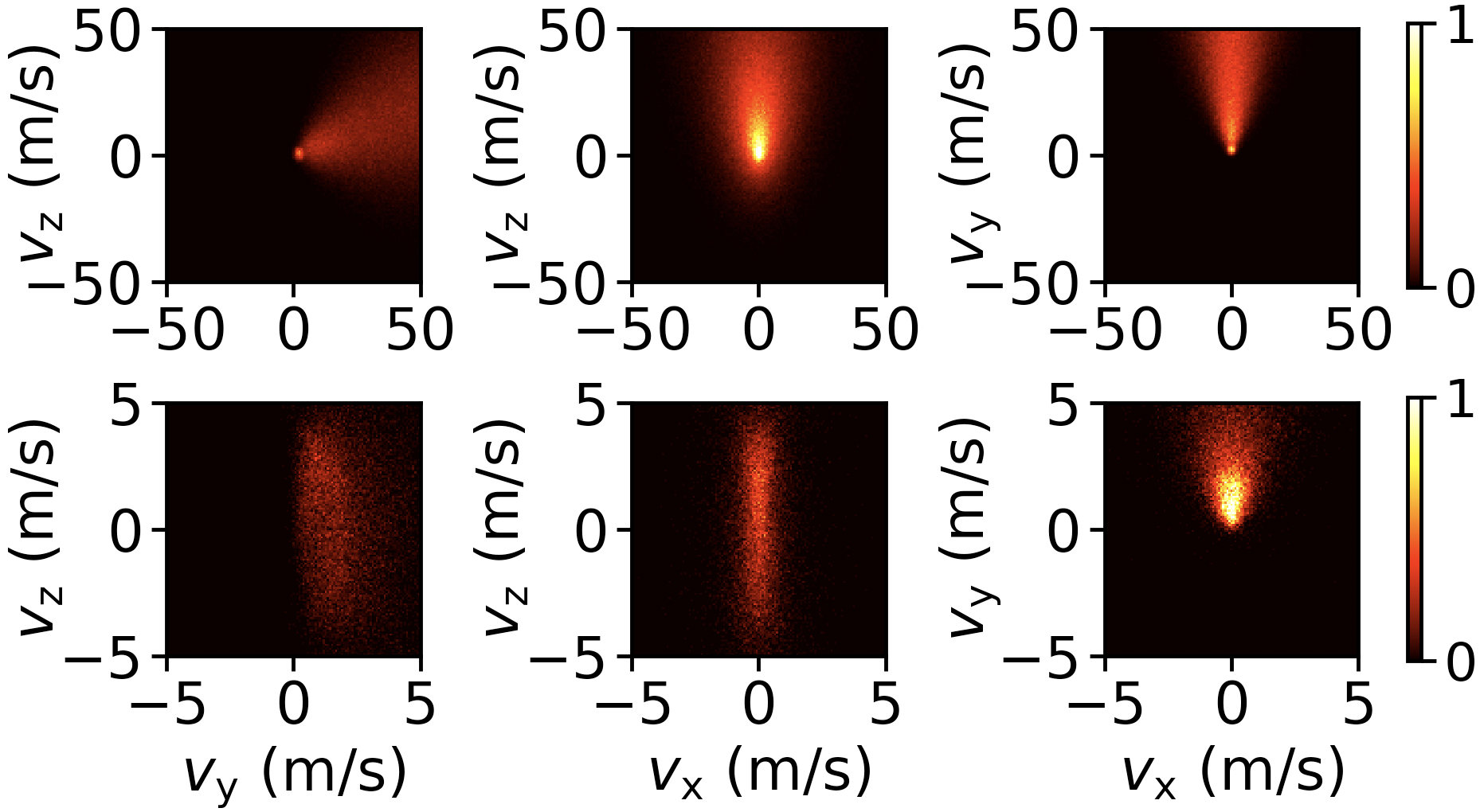}
    \includegraphics[width=\columnwidth]{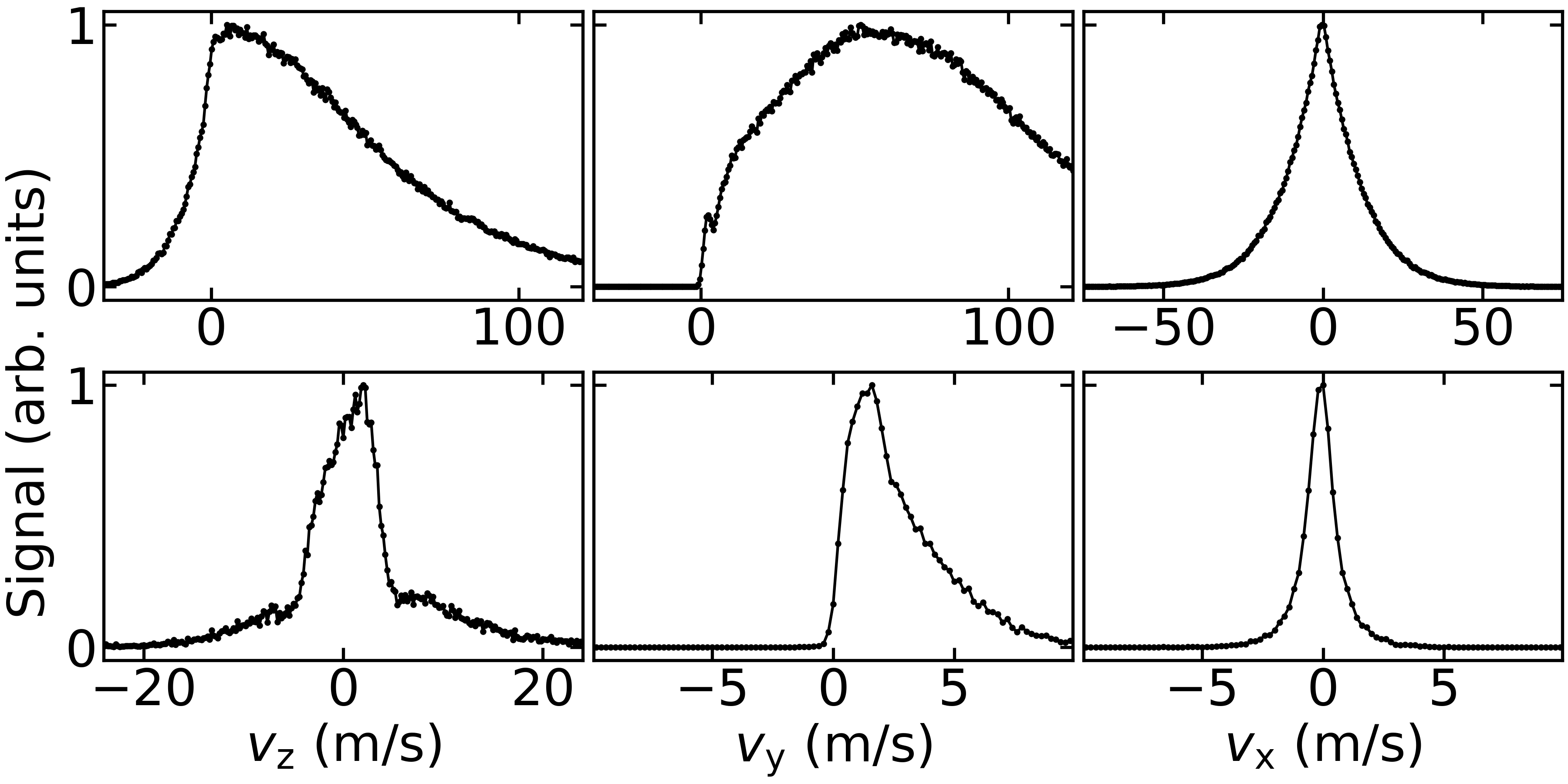}
    \caption{Contour and line graphs of 2D and 1D cuts through  3D $^{87}$Rb velocity distributions 
for two positions along the center line of the jet at a  helium flow rate of 300 sccm.
The velocity distributions are relative to the terminal velocity of the $^4$He atoms at these positions.
We have combined data from all trajectories passing within a 1 mm radius circle perpendicular to the center line. The top row is for a distance of 2 cm from the sonic nozzle while the bottom row is for a distance of 10 cm. The top row of panels shows velocity distributions taken along the center line of the helium jet at a distance of 2 cm from the nozzle, while the bottom row is at a distance of 10 cm.}
    \label{vel_2cm_10cm_Rb}
\end{figure}

Under these conditions, optimal seeding efficiency is found when the collisional thickness is increased by changing the seeding position to 1 cm and increasing the helium flow rate to 300 sccm \footnote{It is possible that a shorter seeding position results in better performance but we have limited the seeding distance due to practical constraints of the beam source geometry}. This results in considerably more collisions for rubidium as compared to lithium as shown in Fig. \ref{collision_comparison}. In Fig.~\ref{Rb_cooling}, the cooling process of rubidium injected into the helium jet is shown. Additionally, the position of all particles that pass the xy plane 4.1 cm in front of the nozzle are shown in Fig.~\ref{simulated_particles_Rb}. As with lithium, a large distribution of kinetic energies is observed far from the nozzle. However, as the cooling process is slower with respect to distance from the nozzle, a clear central peak at $k\times$mK energies is not observed. This is further reflected in the velocity distributions and velocity phase space plots given by Fig.~\ref{vel_2cm_10cm_Rb} where the asymmetry in $v_{\rm z}$ and $v_{\rm y}$ prevents reliable fitting to a central peak. The qualitative difference in the shape of the distributions as compared to lithium can be in part understood due to the lower initial speed of the rubidium atoms. The mean speed for a rubidium source at 550 K is 435 m/s, compared to 1800 m/s for a lithium source at 800 K. A rubidium atom that is entrained and partially thermalized along the center-line of the jet can remain along the center-line longer than a lithium atom with a higher velocity. 

Differences in the velocity distributions of $v_{\rm x}$, $v_{\rm y}$, and $v_{\rm z}$ can be understood by considering geometric constraints. The velocity distributions are analyzed over a 1 mm radius circle centered at $(x, y) = (0, 0)$. As the atoms are injected from an oven beneath the jet centered at $x = 0$, this sets a limit on the maximum value of $v_{\rm x}$ for a given initial speed. The same constraint does not apply for $v_{\rm z}$. It is interesting that the distribution for $v_{\rm y}$ has a sharp cutoff compared to the other two velocity profiles. All rubidium atoms initially have a large $v_{\rm y}$ in the moving frame. Since the mass of $^{87}$Rb is $\approx$ 20 times larger than that of $^4$He, this ensures that a $^{87}$Rb atom deflected downwards has essentially thermalized with the helium. This results in the apparent sharp cutoff and asymmetry as there is a larger distribution of partially thermalized rubidium atoms. 

\begin{figure}[t]
    \centering
 
    \includegraphics[width= 0.85\columnwidth]{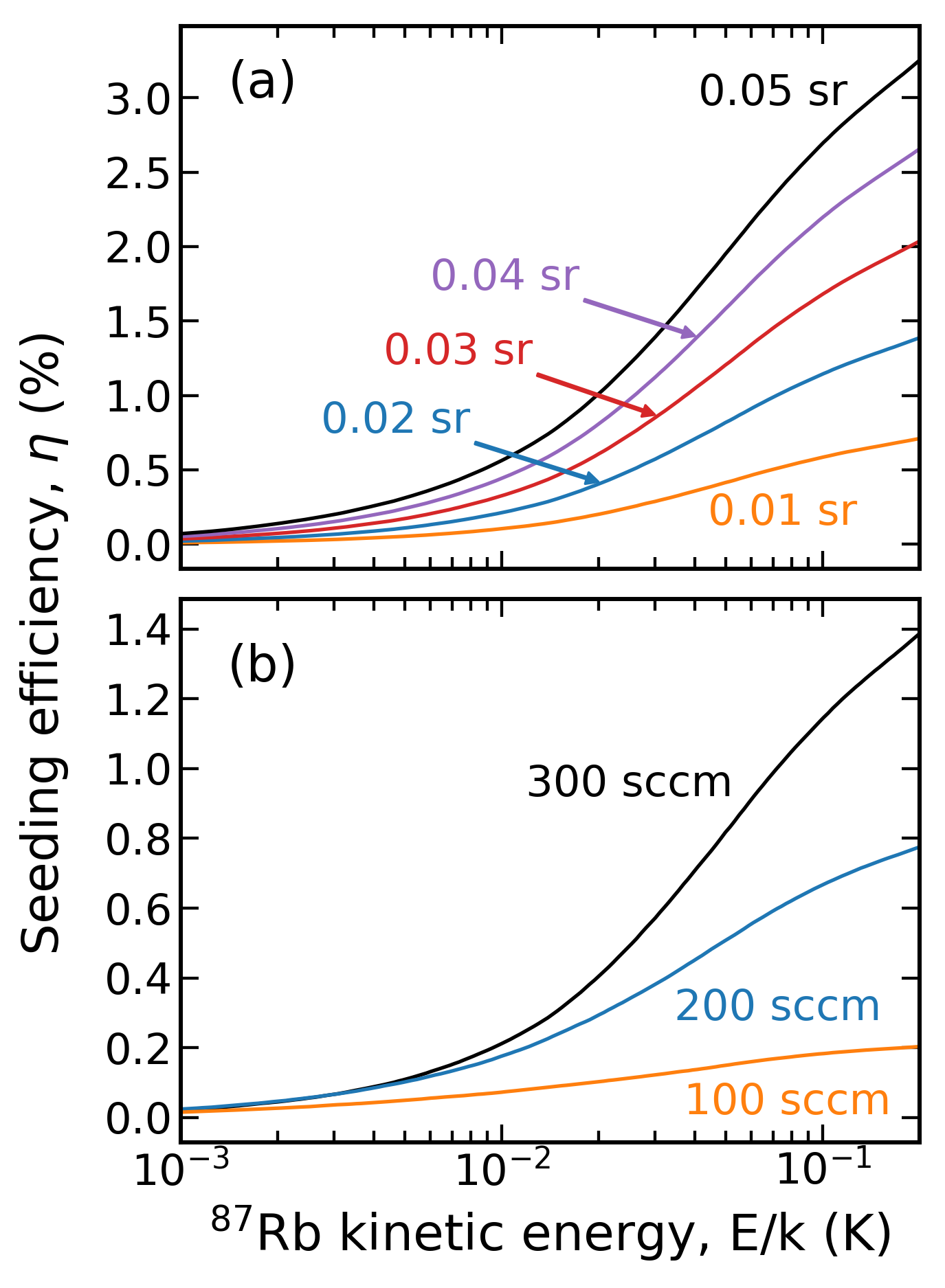}
    \caption{Percentage of simulated $^{87}$Rb atoms with energy in the moving frame up to the value specified on the x-axis. Results are at a distance of 10 cm from the sonic nozzle with the solid angle defined relative to the location of the sonic nozzle. (a) Results for various solid angles with a helium flow rate of 300 sccm. (b) Results for various helium flow rates and a solid angle of 0.02 sr.}
    \label{seeding_efficiency_Rb}  
    
\end{figure}

The seeding efficiency for rubidium for a variety of flow rates and solid angles is shown in Fig.~\ref{seeding_efficiency_Rb} for a distance of $L=10$ cm from the nozzle. 
If a velocity filter was used to remove faster moving rubidium, then it is predicted that at 300 sccm of helium flow, $\approx 1\%$ of the seeded rubidium would leave the skimmer with energies less than 75 mK in the moving frame of the helium.

\section{Conclusion}

We have simulated post-nozzle seeding into a supersonic helium jet. Conditions that maximize seeded flux agree with experimental measurements as well as provide reasonable agreement to measured spectral profiles. Results show that captured lithium and rubidium atoms have energies in the moving frame of the jet on the order of single to tens of $k\times$mK. Since efficient seeding of rubidium is indicated, it may be possible to seed other heavier species such as molecules. 
To achieve efficient seeding the species would need to have collision cross sections with helium which increase drastically at low collision energies. All of the alkali atoms have these properties but it is less clear if this is true for other species. If so, then this method could be used to form an intense beam of cold atoms and molecules by extracting the seeded species with a magnetic or electrostatic lens \cite{Huntington_2022_Cold_Atom_Source, Meijer_1999_PolarMolecules}. Such beams can be useful for studies of cold collisions, atom optics, and precision measurements. Alternatively, cold collision studies could be carried out in the region close to the nozzle.

The simplified model of the jet, while not capturing all relevant features, provides a suitable starting point for understanding seeding dynamics. A 3D Direct Simulation Monte Carlo (DSMC) approach that simulates the helium jet and lithium beam would accurately capture the jet dynamics but would be significantly more computationally expensive \cite{Usami_1999_DSMC}. It would however allow for a detailed study of heating dynamics. Additionally, adding inelastic collisions to the model would allow for the study of rotational cooling of seeded molecules \cite{Schullian_2015_inelastic_collisions_DSMC, Doppelbauer_2017_DSMC_buffer_gas}. If efficient rotational cooling is possible, then our approach could enable  cooling similar to what is achieved in buffer gas-based sources but with considerably better translational cooling \cite{Doyle_2012_buffer_gas_sources}.

\section{Acknowledgements}

We gratefully acknowledge the financial support of this work by Fondren Foundation and the Army Research Laboratory Cooperative Research and Development Agreement number 16-080-004.

\section{Appendix}

\begin{figure}[t]
    \includegraphics[width=0.9\columnwidth,trim=10 0 0 0,clip]{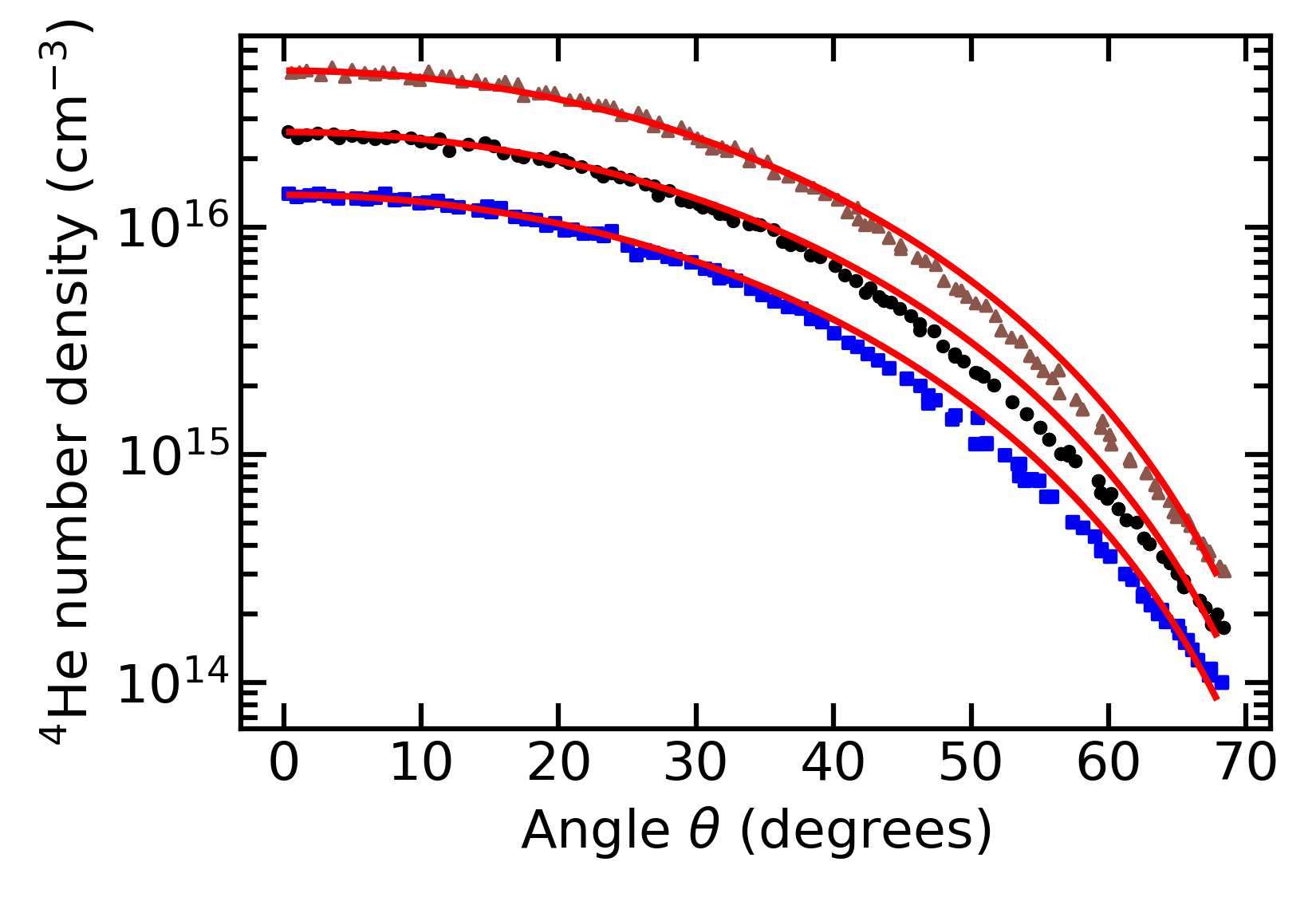}
    \caption{Angular number density profile of a $^4$He expansion with simulated results (markers) and analytic profile (red curve) as a function of angle $\theta$. Results are for $^4$He densities at the nozzle opening with diameter $d_{\rm N} = 0.2$ mm and nozzle opening temperatures of $4\times10^{19}$ cm$^{-3}$ and 300 K (brown triangles), $2\times10^{19}$ cm$^{-3}$ and 100 K (black circles), and $1\times10^{19}$ cm$^{-3}$ and 200 K (blue squares).}
    \label{jet_density_profile}
\end{figure}

We have compared our static model for the helium jet to direct simulation Monte Carlo (DSMC) calculations of helium expansions. These are performed using the DS2V software, a direct simulation Monte Carlo program written by G. Bird \cite{DS2V}. For the computational resources available to us, simulations of the helium expansion are limited to source conditions which correspond to substantially lower collision rates than what is expected in our system. Furthermore, the simulations begin at the opening of the nozzle where the source conditions such as temperature, density, and velocity are specified at the nozzle opening and not equal to the stagnation conditions well inside the nozzle reservoir. Nevertheless, the collision rate and mean free path are such that the flow begins well into the continuum regime. The criteria for continuum flow can be described using the Knudsen number (Kn), given by
\begin{equation}
    \text{Kn} = \frac{\lambda_{\rm mfp}}{L} \,,
\end{equation}
where $\lambda_{\rm mfp}$ is the mean free path and $L$ is a characteristic length scale. Generally, the condition for continuum flow is satisfied when $\text{Kn} < 0.01$ \cite{Bird_1994_DSMC}. The Knudsen numbers for simulations of the helium jet at the opening of the nozzle, where $L$ is the nozzle diameter, are such that $\text{Kn} < 0.005$. 

For a given nozzle shape and opening nozzle diameter, the angular density profile is expected to be independent of the stagnation density and temperature so long as the flow begins in the continuum regime. As such, the helium expansion can be simulated to large angles from the centerline to compare with Eq.~(\ref{eq:fangular}). Results comparing the analytic profile to simulations for various nozzle temperatures and densities are shown in Fig.~\ref{jet_density_profile}. Agreement with Eq.~(\ref{eq:HeDensity}) is generally within $6\%$ for angles $\theta$ up to 30 degrees with inaccuracies of up to $30\%$ at larger angles around 55 degrees.

\begin{figure}[t]
    \includegraphics[width=0.97\columnwidth,trim=10 0 0 0,clip]{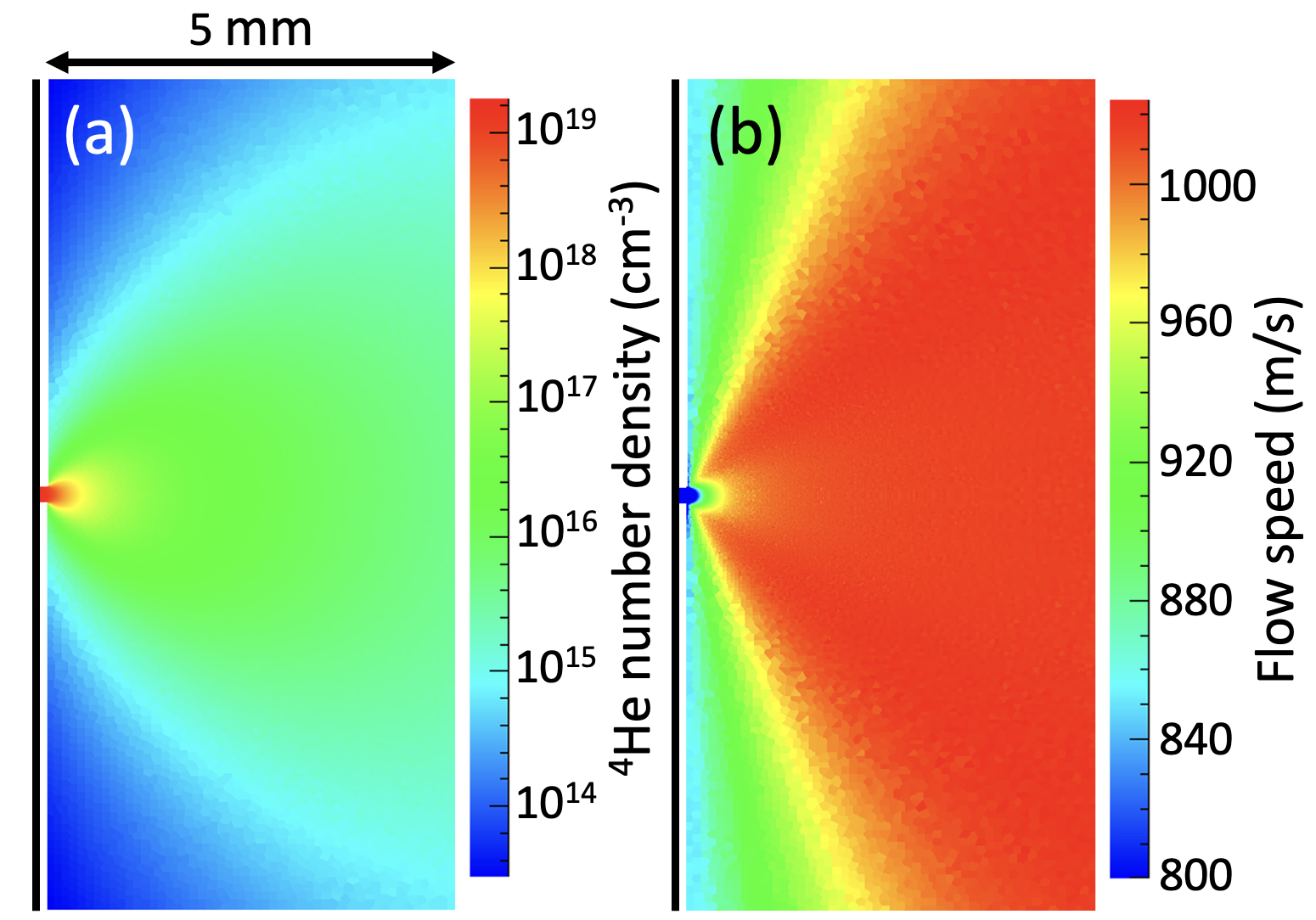}
    \caption{Simulated $^4$He number density profile (a) and flow speed profile (b) for a helium expansion with a nozzle opening temperature of 100 K and nozzle diameter of 0.2 mm.}
    \label{DSMC_results}
\end{figure}

Density and speed profiles are presented in Fig.~\ref{DSMC_results} for a sonic nozzle with an opening diameter of 0.2 mm. The flow speed decreases from the expected terminal velocity for large angles with the speed dropping by $5\%$ from $\theta = 0$ to $\theta = 70$ degrees. At large angles the collision rate is not sufficient to maintain local thermal equilibrium and the conversion of random thermal motion into forward kinetic energy ceases after a short distance. As our source has a greater collision rate, the variations in flow speed in Fig.~\ref{DSMC_results} are expected to be an upper bound for variations we might expect. Nevertheless, the disagreement with Eq.~(\ref{eq:vtv}) at large angles is negligible as the AM-$^4$He relative velocities are dominated by the AM velocity at large angles. 

\begin{figure}[t]
\includegraphics[width=0.97\columnwidth]{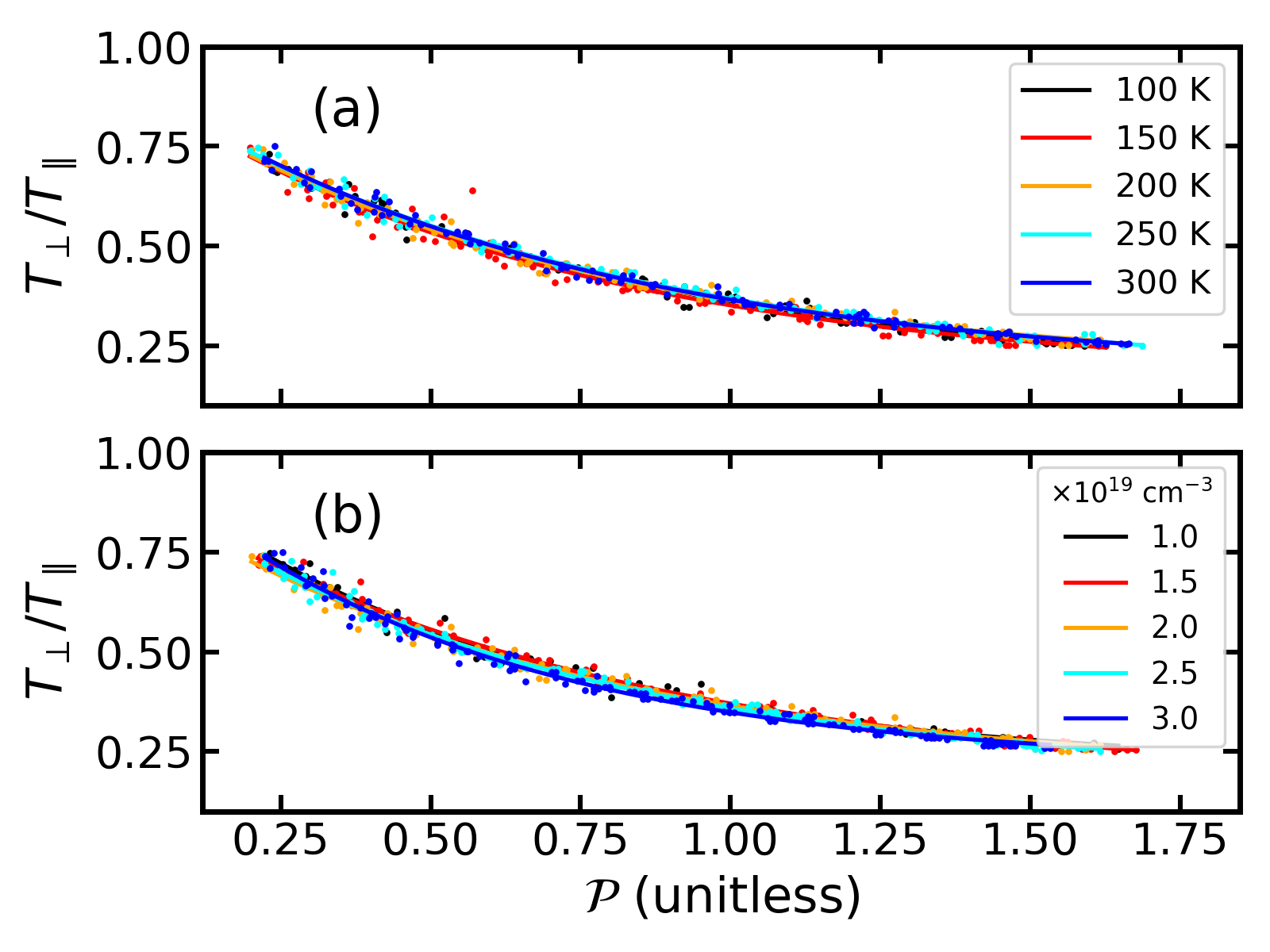}
    \caption{Temperature ratio plotted as a function of the non-equilibrium parameter $\cal P$ for various conditions at the opening of the nozzle. DSMC simulated results (markers) are fit to an inverse power law (solid lines). (a) Nozzle opening density fixed at $2\times 10^{19}$ cm$^{-3}$ with varying starting temperatures. (b) Nozzle opening temperature fixed at $200$ K with varying nozzle opening densities.}
    \label{fig:T_ratio}
\end{figure}

\begin{figure}[t]
\includegraphics[width=0.97\columnwidth,trim=10 0 0 0,clip]{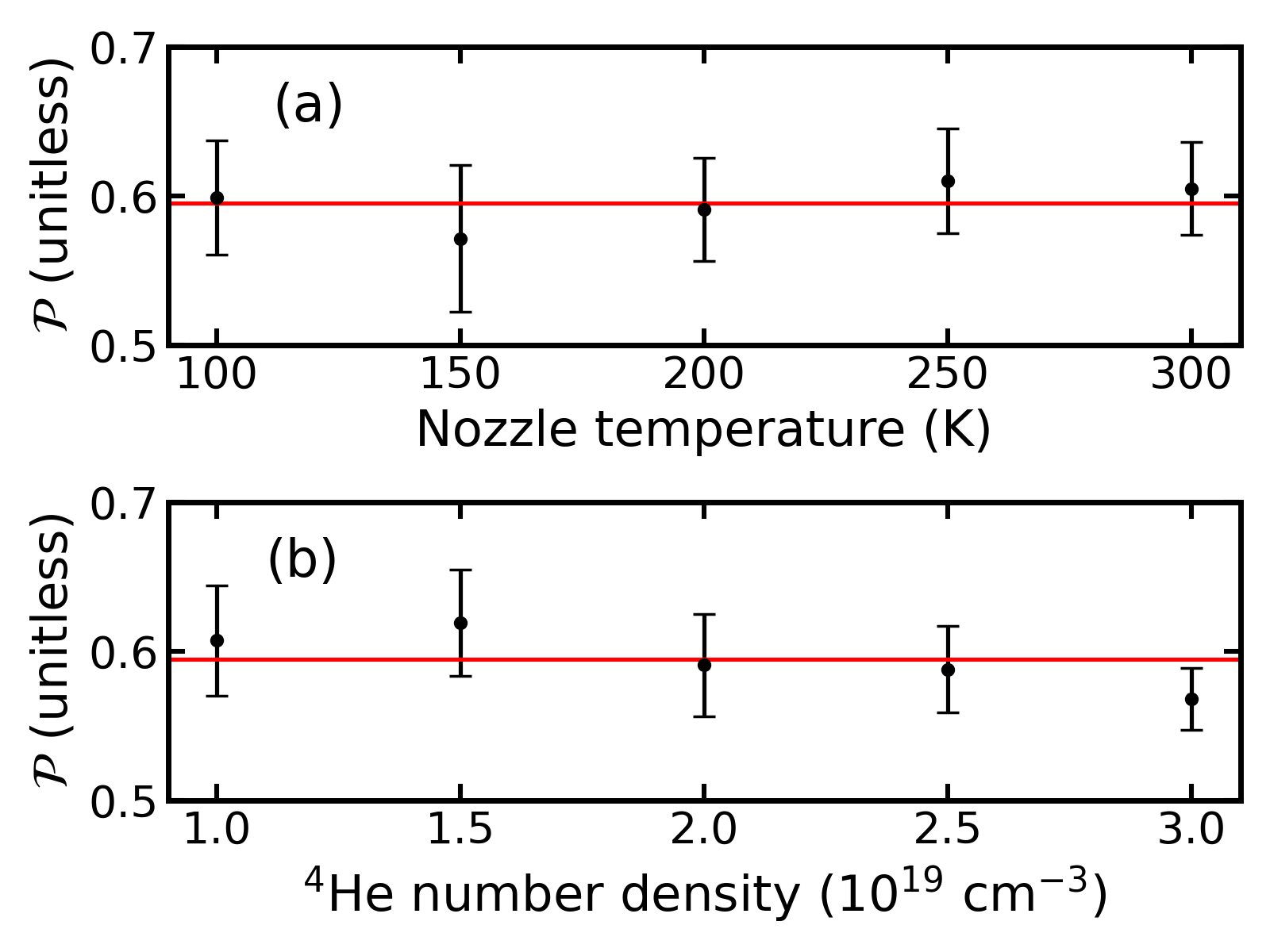}
    \caption{Value of the non-equilibrium parameter $\cal P$ for when $T_{\perp}/T_{\parallel} = 0.5$ for various starting conditions in the DSMC simulation. Results for a nozzle opening density of $2\times10^{19}$ cm$^{-3}$ with varying nozzle opening temperature are show in (a) while varying nozzle opening densities for a nozzle opening temperature of 200 K are shown in (b).}
    \label{fig:P_residuals}
\end{figure}

\begin{figure}[t]
\includegraphics[width=0.97\columnwidth,trim=10 0 0 0,clip]{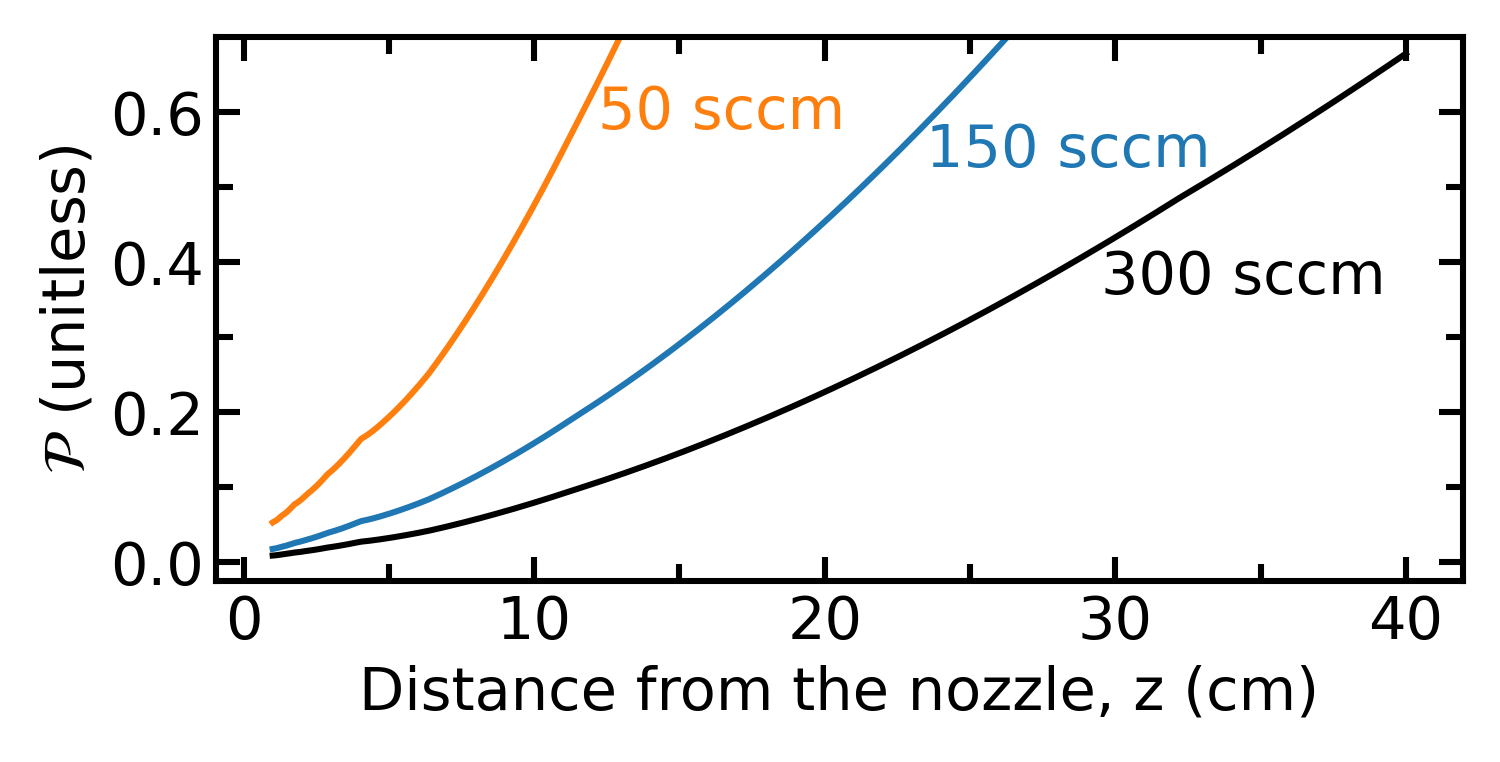}
    \caption{$\cal P$ parameter versus distance from the helium nozzle for various helium flow rates. The $\cal P$ parameter is calculated assuming an adiabatic expansion of the helium.}
    \label{fig:P_vs_z}
\end{figure}

In the lithium seeding simulation, the calculation of the helium jet parameters relies on the assumption that our region of interest is within the continuum-flow regime of the jet. To examine the validity of this assumption, we investigated the transition of a helium expansion from the continuum-flow regime to the free-molecular flow regime using the DS2V software. The ratio $T_\perp/T_\parallel$ of transverse to longitudinal temperature along the centerline of the jet is commonly used as a parameter to quantify the level of thermal equilibrium as the jet transitions from continuum to free-molecular flow \cite{Toennies_1977_theory, Beijerinck_1981_perp_temp}. It has been shown \cite{Bird_1994_DSMC} that the continuum flow assumptions are valid if the collision rate $\Gamma$ is much larger than the proportional rate of change of density $n_{\rm{He}}$, such that 
\begin{equation}
    \Gamma \gg \frac{1}{n_{\rm{He}}}\frac{\text{d}n_{\rm{He}}}{\text{d}t} \,.
\end{equation}
This motivated \cite{Bird_1970_BreakdownParameter} the definition of a non-equilibrium parameter $\cal P$ defined as
\begin{equation}
    {\cal P} = \frac{1}{\Gamma}\left|\frac{\text{d}(\ln n_{\rm{He}})}{\text{d}t}\right| \,,
    \label{eq:P_def}
\end{equation}
which serves as an approximate universal non-equilibrium parameter when investigating the behavior of $T_\perp/T_\parallel$ for various jet conditions \cite{Zarvin_1981_P_parameter, Boyd_1995_P_parameter}. As we are interested in near-continuum flow, Eq.~(\ref{eq:P_def}) can be expressed as
\begin{equation}
  {\cal P} = \frac{v_{\rm s} \lambda_{\rm{mfp}}}{\overline vn_{\rm{He}}}\left|\frac{\text{d}n_{\rm{He}}}{\text{d}r}\right|  \,,
\end{equation}
where $v_{\rm s}$ is the flow speed in the lab frame and $\overline{v}$ is the mean velocity in the moving frame. Recognizing from Eq.~(\ref{eq:HeDensity}) that 
\begin{equation}
    \frac{1}{n_{\rm He}} \left|\frac{\text{d}n_{\rm He}}{\text{d}r} \right|  = \frac{2}{r}\,,
\end{equation}
$\cal{P}$ can now be defined as 
\begin{equation}
    {\cal P} = \frac{2v_s \lambda_{\rm mfp}}{\overline{v}r} \,.
    \label{eq:P_DS2V}
\end{equation}
Using the DS2V software, $\cal P$ is computed using Eq.~(\ref{eq:P_DS2V}) for various densities and temperatures at the opening of the nozzle. The transverse and longitudinal velocity distributions, and thus $T_{\perp}$ and $T_{\parallel}$, are also computed. Plots of $T_{\perp}/T_{\parallel}$ vs $\cal P$ are shown in Fig.~\ref{fig:T_ratio}. We consider the location where $T_{\perp}/T_{\parallel}=0.5$ as a reasonable estimate for when the probability distribution of the expansion can no longer be well approximated by Eq.~(\ref{eq:PHelium}). This occurs when ${\cal P} = 0.60(2)$ as shown in Fig.~\ref{fig:P_residuals}. 

\begin{figure}[t]
\includegraphics[width=0.85\columnwidth,trim=0 0 0 0,clip]{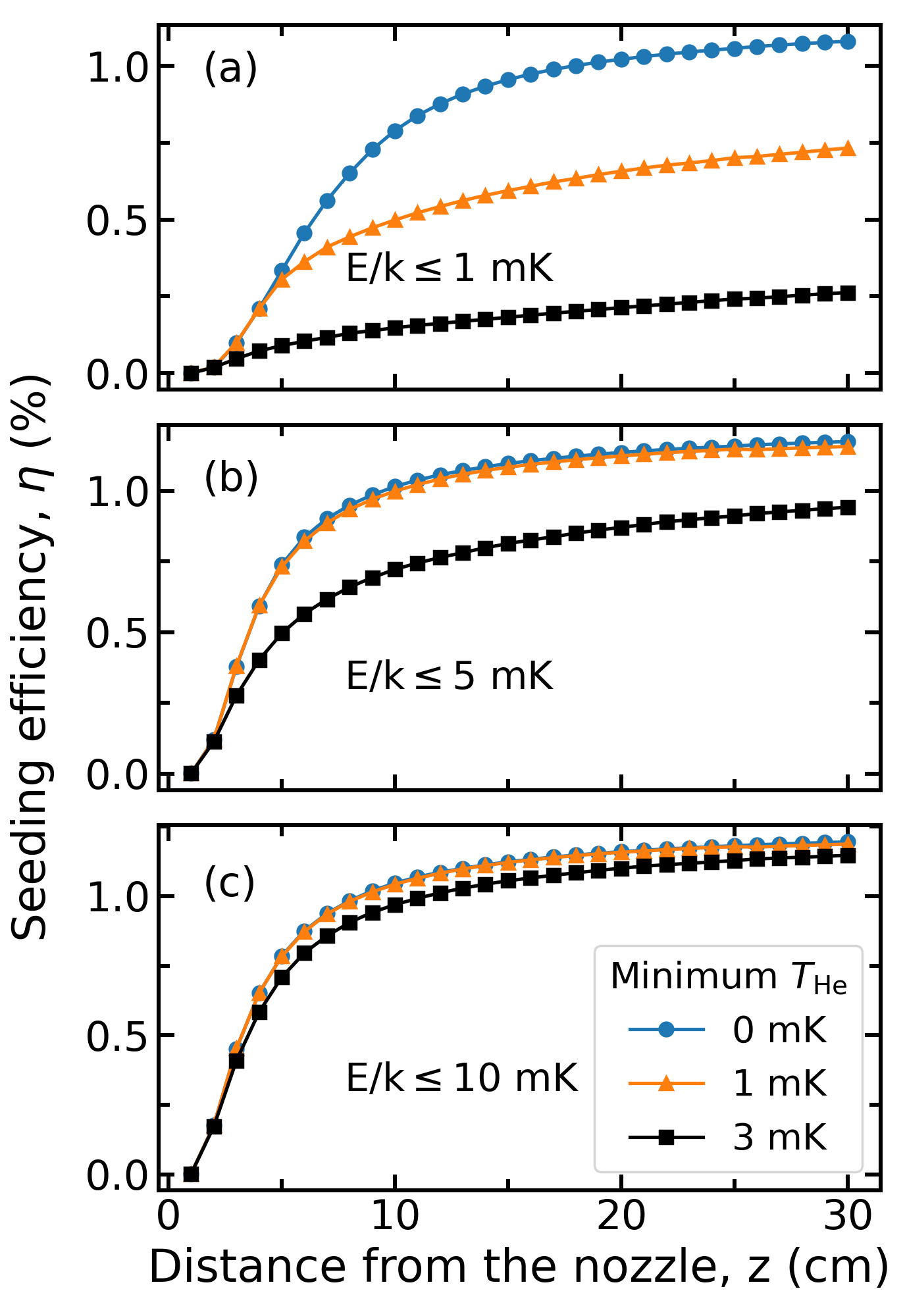}
    \caption{Seeding efficiency of $^7$Li atoms with energies in the moving frame up to and including (a) $k\times1$ mK, (b) $k\times5$ mK, and (c) $k\times10$ mK versus distance from the nozzle. For each subplot, the seeding efficiency is shown for three models of the helium expansion where the minimum helium temperature is set to 0 mK, 1 mK, and 3 mK. The seeding efficiency is specified for a solid angle of 0.02 sr and a helium flow rate of 200 sccm.}
    \label{fig:Li_min_He_plots}
\end{figure}

From Eq.~(\ref{eq:Thelium}), (\ref{eq:vtv}), and (\ref{eq:HeDensity}), along with the $^4$He-$^4$He cross-section from Ref.~\cite{Chrysos_2017_He_He_cross_section}, the parameter $\cal{P}$ can be calculated for our source conditions as a function of distance from the helium nozzle. Results for $\cal{P}$ along the center-line of the expansion for various helium flow rates are shown in Fig.~\ref{fig:P_vs_z}. The values are calculated assuming an adiabatic expansion and will become inaccurate as the value of $\cal{P}$ becomes large. However, at small to intermediate values of ${\cal P}$, the approximate location where the helium collisions begin to turn off can be estimated. Examining the location where ${\cal P} = 0.6$, it can be seen that the jet is expected to remain collisional well past the region where seeding occurs. In fact, at the highest flow rate of 300 sccm, the helium is expected to continue to cool past 20 cm from the nozzle. This would amount to a temperature on the order of 100 $\mu$K. It is unlikely, however, that these temperatures will be reached due to cluster formation. At high helium flow rates, cluster formation  of the helium can occur releasing heat of condensation, setting a limit on the cooling. Some of us describe in Ref.~\cite{Huntington_2022_Cold_Atom_Source} that at our higher helium flow rates, we are likely in a regime where some cluster formation is occurring. While direct observation of clustering in our jet has not been observed, it has been observed in prior experiments with pure helium expansions. In these experiments, the lowest temperatures that have been achieved are typically around 0.5 mK to 1 mK \cite{Wang_1988_mK_He_jet, Luo_1993_HeJet, Ludwig_2002_HeJet, Pedemonte_2003_HeJet}. As a result, it seems reasonable to conclude that the minimum temperature of the $^4$He jet without the heat load of the alkali atoms is likely comparable to past works.

\begin{figure}[t]
\includegraphics[width=0.85\columnwidth,trim=0 0 0 0,clip]{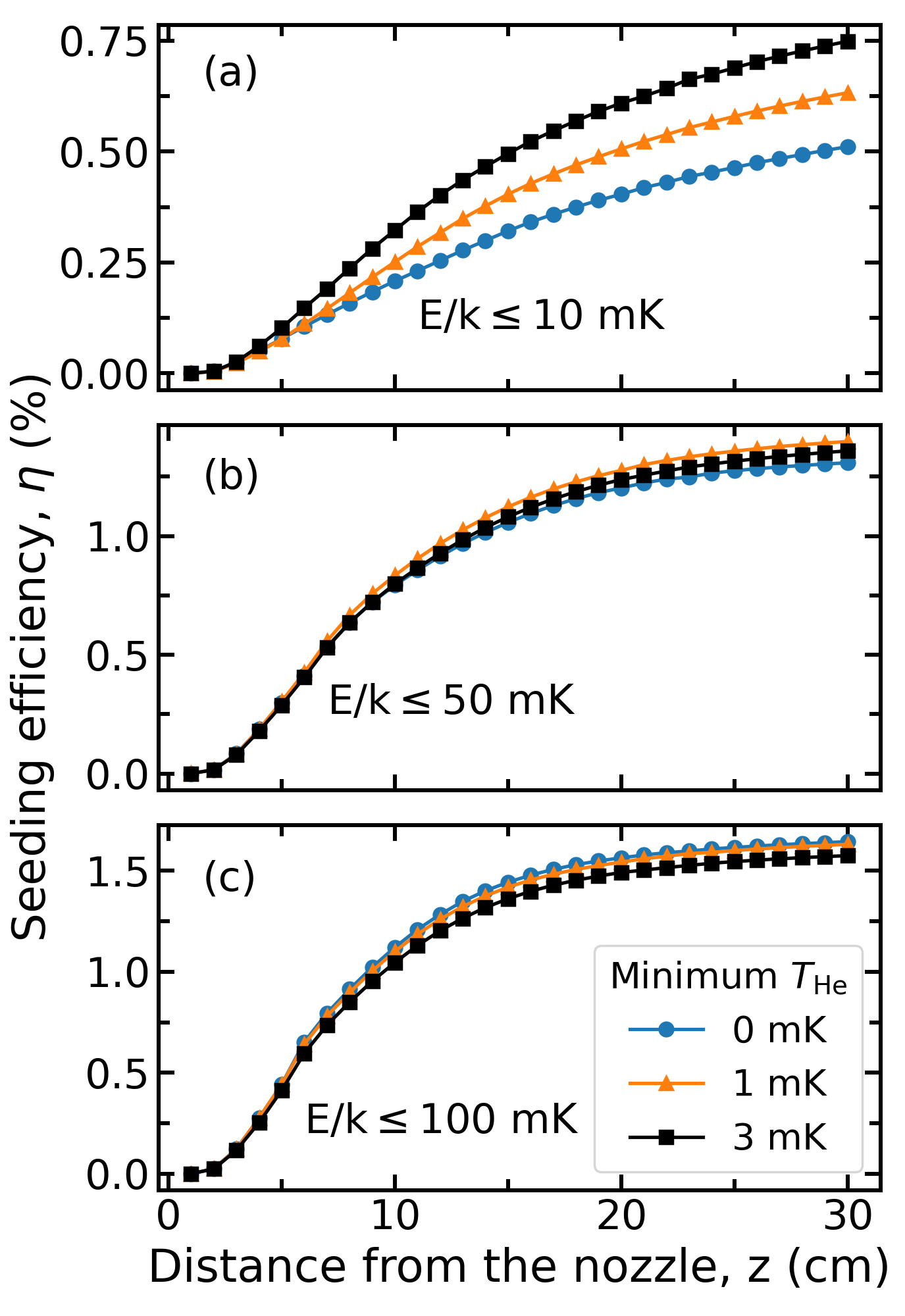}
    \caption{Seeding efficiency of $^{87}$Rb atoms with energies in the moving frame up to and including (a) $k\times10$ mK, (b) $k\times50$ mK, and (c)$k\times100$ mK versus distance from the nozzle. For each subplot, the seeding efficiency is shown for three models of the helium expansion where the minimum helium temperature is set to 0 mK, 1 mK, and 3 mK. The seeding efficiency is specified for a solid angle of 0.02 sr and a helium flow rate of 300 sccm.}
    \label{fig:Rb_min_He_plots}
\end{figure}

While cluster formation is not included in the simulation to determine the ultimate limit on the helium temperature, it is still desirable to explore how variations in the helium temperature affect the seeding efficiency. To do this, an artificial temperature floor in the expansion is introduced such that the helium temperature decreases according to Eq.~(\ref{eq:Thelium}) until it reaches a  minimum value after which the temperature is constant. This is not a physically accurate way to model heating from cluster formation or decoupling as the jet transitions from continuum to free molecular flow. It does, however, allow us to explore with a simple model how modifications to the helium velocity distributions affect seeding efficiency. Three different minimum temperatures of 0 mK, 1 mK, and 3 mK are used in the simulations. Results for the seeding efficiency versus distance from the nozzle for these minimum temperatures are shown in Fig.~\ref{fig:Li_min_He_plots} and Fig~\ref{fig:Rb_min_He_plots}. Here, the seeding efficiency is displayed as the percentage of injected alkali-metal atoms within a solid angle of 0.02 sr and with energies up to and including $k\times1$ mK $k\times5$ mK and $k\times10$ mK for $^7$Li and $k\times 10$ mK, $k\times50$ mK and $k\times 100$ mK for $^{87}$Rb. Unsurprisingly, variations in the helium temperature only have a substantial impact on alkali-metal atoms with low energies in the moving frame. Since this occurs after entrainment, the source conditions that maximize seeding efficiency are unaffected by these variations in the helium temperature.  Interestingly, a higher helium temperature results in a high seeding efficiency for $^{87}$Rb at low energies. A consequence of the $^4$He-$^{87}$Rb resonance at around 10 mK. Overall, it is likely that the simulation does not accurately predict the final seeding efficiency of $^7$Li atoms with energies $\leq k\times1$ mK or $^{87}$Rb atoms with energies $\leq k\times10$ mK. However, it should be reasonably accurate for predicting energies $\geq k\times10$ mK and $\geq k\times100$ mK for $^7$Li and $^{87}$Rb respectively. 

\bibliography{references}

\end{document}